\documentclass[12pt,preprint]{aastex}

\slugcomment{To appear in the Oct 2004 issue of {\it The Astronomical Journal}}

\shorttitle{Five Companions to Nearby Dwarfs}
\shortauthors{Golimowski et al.}
 
\begin{document}
 
\title{The Solar Neighborhood IX: \\ 
{\it Hubble Space Telescope} Detections of Companions to \\ 
Five M and L Dwarfs within 10~pc of the Sun}

\author{David~A.\ Golimowski,\altaffilmark{1}
Todd~J.\ Henry,\altaffilmark{2}
John~E.\ Krist,\altaffilmark{3}
Sergio Dieterich,\altaffilmark{1} \\
Holland~C.\ Ford,\altaffilmark{1}
Garth~D.\ Illingworth,\altaffilmark{4}
David~R.\ Ardila,\altaffilmark{1}
Mark Clampin,\altaffilmark{5} \\
Otto~G.\ Franz,\altaffilmark{6}
Lawrence~H.\ Wasserman,\altaffilmark{6}
G.~Fritz Benedict,\altaffilmark{7}
Barbara~E.\ McArthur,\altaffilmark{7}\\
and
Edmund~G.\ Nelan\altaffilmark{3}
}

\altaffiltext{1}{ 
Department of Physics and Astronomy, 
Johns Hopkins University, 
3400 North Charles Street, 
Baltimore, MD 21218-2686
}
\altaffiltext{2}{ 
Department of Physics and Astronomy, 
Georgia State University, 
Atlanta, GA 30303-3083 
}
\altaffiltext{3}{ 
Space Telescope Science Institute,
3700 San Martin Drive,
Baltimore, MD 21218 
}
\altaffiltext{4}{ 
Lick Observatory,
University of California at Santa Cruz,
1156 High Street,
Santa Cruz, CA 95064
}
\altaffiltext{5}{ 
NASA/Goddard Space Flight Center,
Code 681, 
Greenbelt, MD 20771 
}
\altaffiltext{6}{ 
Lowell Observatory,
1400 West Mars Hill Road, 
Flagstaff, AZ 86001 
}
\altaffiltext{7}{ 
McDonald Observatory,
University of Texas, 
Austin, TX 78712-1083 
}

\begin{abstract} 
We report the detections of low-mass companions to five M and L dwarfs within 10~pc of the Sun using the {\it Hubble Space 
Telescope (HST)} Near-Infrared Camera and Multi-Object Spectrometer (NICMOS).  Follow-up observations using the {\it HST}
Advanced Camera for Surveys (ACS) and Fine Guidance Sensor 1r (FGS1r) confirm our NICMOS discoveries of companions to the L4.5
dwarf GJ~1001B (LHS~102B) and the M5 dwarf LHS~224, respectively.  Images obtained with the Astrophysical Research Consortium
3.5~m telescope at Apache Point Observatory verify our discovery of a companion to the M3 dwarf G~239--25.  Our NICMOS images 
confirm the previously suspected duplicity of the M3 dwarfs GJ~54 and GJ~84.  The components of GJ~1001BC and LHS~224AB have 
nearly equal luminosities in all the ACS and/or NICMOS bandpasses in which they were observed.  The magnitudes of GJ~54A and B
in one FGS1r bandpass and four NICMOS bandpasses differ by $\lesssim 1$.  GJ~84B and G~239--25B are $\sim 4$ magnitudes fainter
than their M3 companions in the NICMOS bandpasses.  We estimate from our NICMOS photometry that GJ~84B and G~239--25B have 
spectral types M7 and M8, respectively, and masses near the lower limit for sustained hydrogen burning.  The apparent separations
of GJ~1001BC, GJ~54AB, and LHS~224AB suggest orbital periods less than 5~yr.   These binary dwarfs are ideal candidates for 
continued astrometric monitoring that will quickly yield accurate dynamic masses needed to constrain the mass--luminosity 
relation for low-mass stars and brown dwarfs.
\end{abstract}
 
\section{Introduction}

Despite their proximity, the stars within 10~pc of the Sun remain an incompletely characterized population.  This assessment 
is especially true for very-low-mass (VLM; $M \lesssim 0.20~M_{\odot}$) dwarfs, whose low luminosities ($L \lesssim 
10^{-2.5}~L_{\odot}$) retard comprehensive studies of their intrinsic properties and numerical abundance.   For example, the 
empirical mass--luminosity relation (MLR) for late-M dwarfs is only sparsely sampled \citep{hen90,hen93,hen99,del00}, and its 
extension to the cooler L and T dwarfs is completely uncharted.  Also, the luminosity function (LF) of the nearest VLM stars is 
unsettled, as M dwarfs within 10~pc continue to be identified as companions (\citealt{gol95,giz96,del99,opp01}; this paper) and 
in isolation \citep{hen97,sch01,mcc02,rei03,tee03,ham04}.  Only by exhaustively searching for faint nearby stars and 
closely-separated companions can the MLR and LF be sufficiently well determined to yield a definitive mass function of VLM stars
and brown dwarfs.

Since July 1997, we have conducted a ``snapshot'' search for VLM companions to all known stars within 10~pc using the {\it Hubble 
Space Telescope} ({\it HST}) Near-Infrared Camera and Multi-Object Spectrometer (NICMOS).  Our search spans the circumstellar 
region 0\farcs2--$10''$ from each target star, which corresponds to 2--100~AU for the most distant stars in our sample.  This 
region bridges the 0--10~AU search-spaces of speckle-interferometry and radial-velocity surveys \citep{hen90,lei97,mar98,mar00} 
and the 100--1000~AU search-spaces of deep, near-infrared imaging surveys \citep{skr89,sim96,kir01,wil01,hin02,mz04}.  With ultimate 
detection limits of $M_J \approx 21$ and $M_K \approx 19.5$ at 10~pc \citep{kri98}, our NICMOS search is sensitive to companions
that are at least 10 magnitudes fainter than the empirical end of the hydrogen-burning main sequence \citep{hen93} and at least 6 
magnitudes fainter than the archetypal T dwarf, GJ~229B \citep{mat96,leg02a}.

The targets in our NICMOS snapshot program have been observed sporadically over seven years in order to fill small ($\sim 20$~min) 
gaps in the {\it HST} observing schedule.  To date, single-epoch images of 239 stars in 191 systems have been recorded.  These 
observations represent 72\% of the {\it HST} pointings required to survey all the stars and brown dwarfs within 10~pc identified 
astrometrically before 1 January 2004.  (We treat known systems with separations less than $4''$ as individual {\it HST} targets.)  
In this paper -- the ninth of the ``The Solar Neighborhood'' series produced by the Research Consortium on Nearby Stars (RECONS)
\footnote{Information about RECONS is currently available on the World Wide Web at http://www.chara.gsu.edu/RECONS/.} -- we report 
the discoveries of VLM companions to three nearby M and L dwarfs (G~239--25, GJ~1001B, and LHS~224) and we confirm companions to 
two other M dwarfs (GJ~54 and GJ~84).  All but one of these binary dwarfs have projected separations less than 5~AU.  We discuss
the potential impact of these companions on the MLR, and we update the state of multiplicity among presently known members of 
the RECONS ``10~pc sample.''

\section{Observations and Data Processing}

Table~1 contains basic information about the five M and L dwarfs to which we report newly discovered or confirmed VLM companions.  
Information about GJ~1001A (also known as LHS~102A) is also included to provide a full description of the GJ~1001 system.  Table~2 
is a log of initial and follow-up observations of each dwarf conducted with NICMOS, the {\it HST} Advanced Camera for Surveys (ACS)
and Fine Guidance Sensor 1r (FGS1r), and the Astrophysical Research Consortium (ARC) 3.5~m telescope at Apache Point Observatory.  
We now describe these observations in detail.

\subsection{NICMOS Observations}

\begin{figure}[t]
   \epsscale{1.11}\plottwo{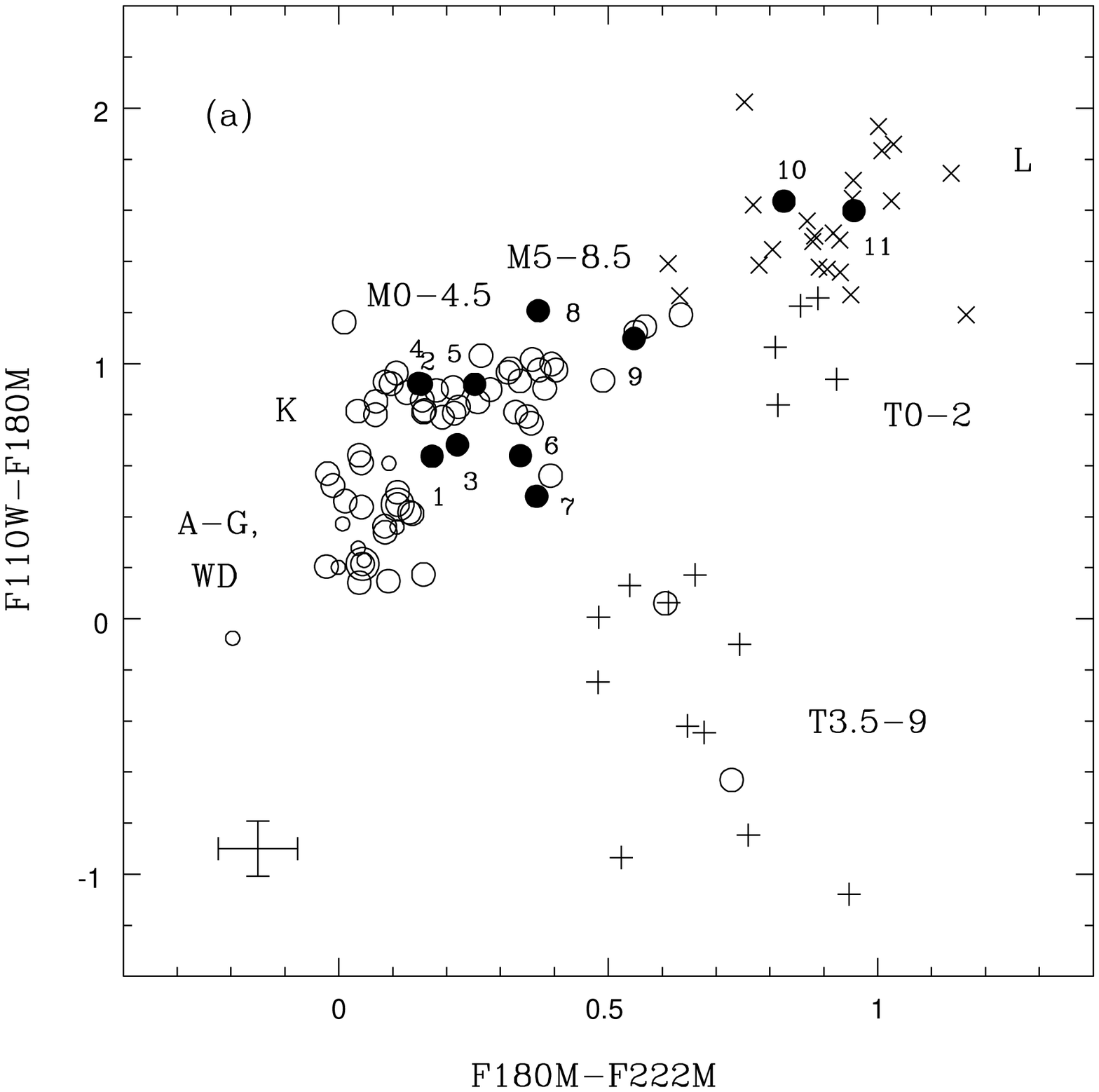}{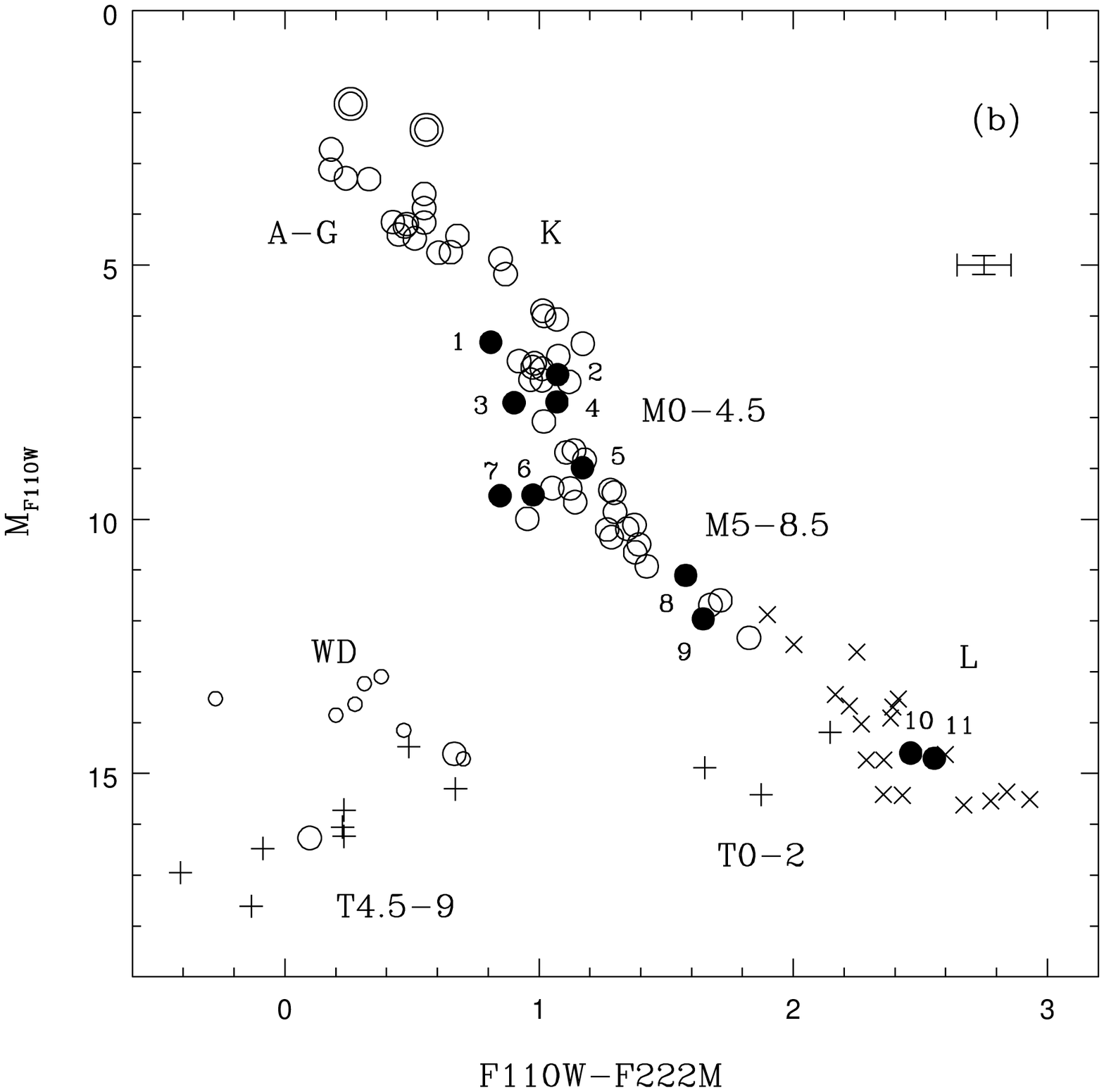}
   \caption{
	(a) NICMOS color--color diagram of F110W--F180M versus F180M--F222M.  (b) NICMOS color--magnitude diagram of $M_{\rm F110W}$ 
	versus F110W--F222M.  In each diagram, open circles represent stars and brown dwarfs within 10~pc imaged in our snaphot survey.
	(Large circles represent dwarfs with spectral types A through T, small circles represent white dwarfs, and concentric circles 
	represent subgiants.)  Filled and enumerated circles depict the following components of systems reported in this paper: 
        (1)~GJ~84A, (2)~GJ~54A, (3)~GJ~54B, (4)~G~239--25A, (5)~GJ~1001A, (6)~LHS~224A, (7)~LHS~224B, (8)~GJ~84B, (9)~G~239--25B, 
        (10)~GJ~1001B, and (11)~GJ~1001C.  Typical measurement errors for these nine dwarfs are shown in the lower-left corner of (a)
        and the upper-right corner of (b).  The $\times$ and $+$ symbols represent synthetic magnitudes and colors computed for L and T
        dwarfs, respectively.  These values were obtained using flux-calibrated, near-infrared spectra \citep{geb02,kna04}, weighted-mean
        trigonometric parallaxes \citep[and references therein]{gol04}, and the NICMOS Exposure Time Calculator produced by the Space 
        Telescope Science Institute.
	}
  \label{Golimowski.fig1}
\end{figure}

Each dwarf was observed at a single epoch with NICMOS Camera~2 (NIC2), which has a pixel scale of 0\farcs076 and a field-of-view 
(FOV) of 19\farcs5~$\times$~19\farcs5 \citep{roy03}.  The dwarfs were acquired as close to the center of the NIC2 field as 
the uncertainties in their coordinates and proper motions allowed.  Images were recorded in fine-guidance mode through the F110W, 
F180M, F207M, and F222M filters (Table~3), which alternately sample the 1--$2.5~\mu$m spectra of putative brown-dwarf companions 
outside and inside the CH$_4$ absorption bands at 1.6--$1.8~\mu$m and 2.2--$2.4~\mu$m.  Figures~\ref{Golimowski.fig1}a and \ref{Golimowski.fig1}b 
show that this combination of filters permits easy discrimination of late-M, L, and T dwarfs from most background stars.  We use 
this color discrimination and expectations based on Galactic latitude to identify candidate VLM companions among the many faint 
sources often imaged in a target field.

\begin{figure}[t]
   \epsscale{1}\plotone{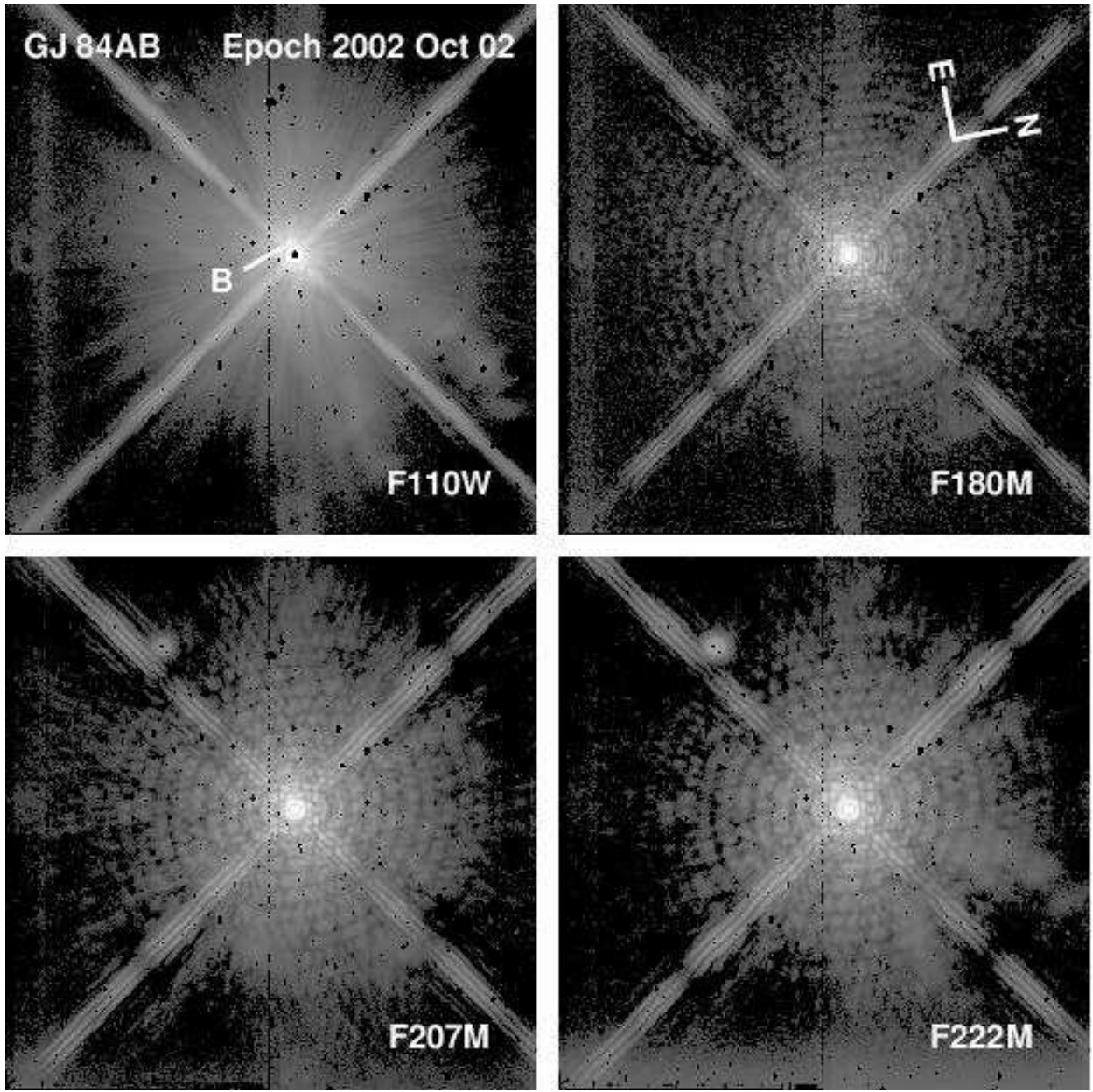}
   \caption{
       Reduced and calibrated NIC2 images of the binary M dwarf GJ~84AB.  Each panel shows the full NIC2 FOV 
       (19\farcs5~$\times$~19\farcs5) with logarithmic pixel scaling.  The saturated core of the F110W image is 
       marked with black pixels.  The other black pixels denote intrinsically bad pixels and pixels affected by
       instrumental debris (``grot'') that has settled on the detector.   The diffuse disks in the upper left 
       quadrants of the images are flat-field artifacts caused by NIC2's coronagraphic hole.  The M7~{\small V}
       companion {\it (marked ``B'' in the upper left panel)}, which lies 0\farcs44 from the primary star along the 
       southeast diffraction spike, is barely seen amid the highly structured PSF of the primary star.  
   }
  \label{Golimowski.fig2}
\end{figure}

Images through each filter were recorded using NICMOS's multiple-accumulate (MULTIACCUM) mode, which allows the detector to be read 
nondestructively in predefined sampling sequences.  Two sets of exposures totalling 64~s were recorded through each of the F110W 
and F180M filters, and two sets of exposures totalling 128~s were recorded through each of the F207M and F222M filters.  All sets 
of exposures were recorded using a logarithmic sampling sequence that provides large dynamic range among the constituent images.
The images were reduced using the standard NICMOS calibration software \citep{dic02} and appropriate reference images and tables.
The calibration software combines the constituent images of each sampling sequence so that values of pixels ultimately affected by 
cosmic-ray impacts or saturation are replaced by appropriately scaled values of those pixels sampled before the contaminating event.
Consequently, the lost data in the reduced images are limited to instrinsically bad pixels and/or pixels that were saturated before 
the first nondestructive read of the detector (i.e., within the first 0.303~s of the exposure).  Figure~\ref{Golimowski.fig2} shows the set
of reduced and calibrated images obtained for a typical target, namely, the binary M dwarf GJ~84AB.  

\begin{figure}[t]
   \epsscale{1}\plotone{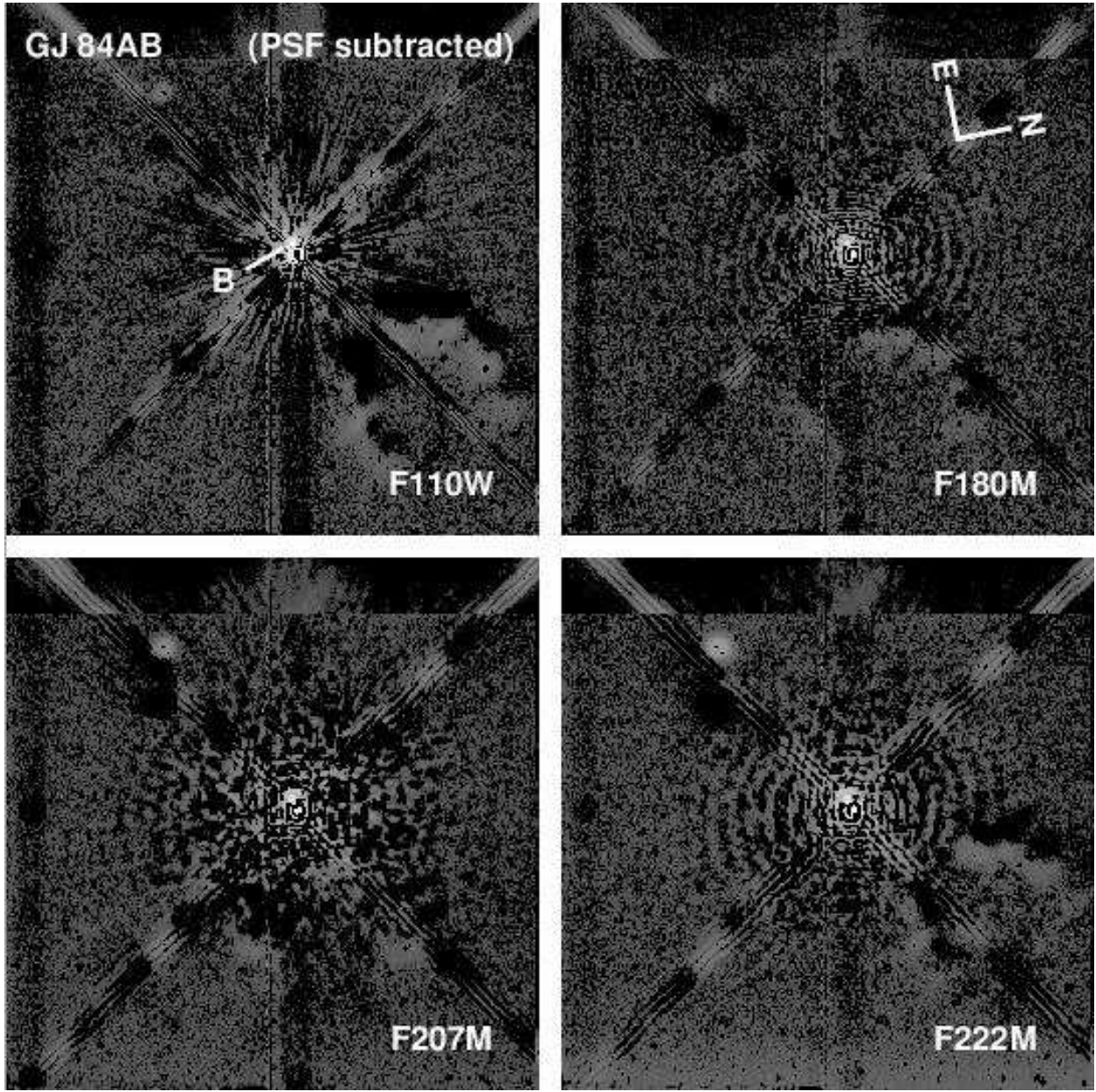}
   \caption{
       NIC2 images of GJ~84AB after optimal shifting, scaling, and subtraction of reference images of GJ~643 
       (M3.5~{\small V}) obtained on 2002~October 15.  Each panel shows the full NIC2 FOV (19\farcs5~$\times$~19\farcs5)
       with logarithmic pixel scaling.  The top 27 rows of each image are unaltered by the PSF subtraction because 
       GJ~643's image was located 27 rows higher than GJ~84's image.  The M7~{\small V} companion {\it (marked ``B'' in the 
       upper left panel)} is evident in each panel immediately to the east of the central residuals of each subtracted PSF.
       No other sources appear in the FOV.
   }
  \label{Golimowski.fig3}
\end{figure}

To facilitate the detection of faint companions in the reduced images, we subtracted the point-spread functions (PSFs) of the target
stars in the manner described by \citet{kri98}.  First, we selected reference PSFs for each target and filter from the large collection
of single-star images recorded during our snapshot survey.  The reference PSFs for a given target were those that best matched not only
the brightness, spectral type, and field position of the target, but also the location of NIC2's migratory cold mask at the time of the 
observation.  The reference PSFs were then iteratively shifted, scaled, and subtracted from the target images until the residuals were 
visually minimized.  Figure~\ref{Golimowski.fig3} shows the results of this technique for the images of GJ~84AB shown in Figure~\ref{Golimowski.fig2}.  
Our imaging and processing strategies yield sensitivities to faint companions that are dependent on the brightness of the target,
the quality of the PSF subtraction, and the field position.  \citet{kri98} reported our detection limits at various angular distances 
from representative targets whose PSF-subtracted images are of average quality.

We measured the apparent magnitudes of unsaturated and well-separated targets and field sources using conventional aperture photometry.
To convert instrumental count rates to Vega-based magnitudes, we used aperture corrections computed for our selection of NICMOS filters
\citep{kri98} and the conversion recipe described in the {\it HST} Data Handbook for NICMOS \citep{dic02}: 

\begin{equation}
m = -2.5~{\rm log} (PHOTFNU \times CR \times [F_{\nu}\rm{(Vega)}]^{-1}) + ZP({\rm Vega}),
\end{equation}

\noindent 
where $PHOTFNU$ is the bandpass-averaged flux density for a source producing a count rate of 1~DN~s$^{-1}$, $CR$ is the count rate 
(in DN~s$^{-1}$) of the observed source measured for an infinite aperture, $F_{\nu}$(Vega) is the flux of a zero magnitude star with a 
Vega-like spectrum, and $ZP({\rm Vega}) = 0.02$ is the magnitude of Vega in all NICMOS bandpasses \citep{cam85}.  Table~3 lists the 
values of $PHOTFNU$ and $F_{\nu}$(Vega) adopted for images obtained before and after the installation of the NICMOS Cooling System 
(NCS) in March 2002.  

We computed the magnitudes of closely-separated point sources from model NIC2 PSFs that were generated using the Tiny Tim software 
package \citep{kri03} and then fitted to each overlapping image.\footnote{We used model PSFs instead of natural reference PSFs because
Tiny Tim produces oversampled images that can be more accurately fitted to the overlapping and undersampled NIC2 images.  The model 
PSFs were generated with five times the pixel resolution of NIC2, aligned and scaled to each imaged source and then binned to the 
nominal NIC2 resolution to assess the quality of the fit.}  To verify the fidelity of this method, we compared the four-band magnitudes
of 35--80 single stars obtained from fitted Tiny Tim PSFs with those of the same stars obtained via conventional aperture photometry.
The fitted F207M and F222M magnitudes were, on average, indistinguishable from those obtained via aperture measurements.  The fitted 
F110W and F180M magnitudes, however, differed from the aperture magnitudes by an average of $+0.013$ and $-0.022$~mag, respectively.
We therefore corrected the F110W and F180M magnitudes of closely-separated sources by subtracting these average offsets.  The 
root-mean-square (RMS) deviations between the model and aperture magnitudes of our single-star sample were $\sim 0.08$~mag for F110W
and only $\sim 0.03$~mag for the other filters.  This consistency encouraged us to estimate the magnitudes of overexposed targets 
using model PSFs fitted to the unsaturated parts of their images.  We investigated this procedure using artificially saturated images
of a well-exposed star.  We found that the fitted PSFs yielded progressively fainter magnitudes as the numbers of saturated pixels 
increased.  Consequently, we adjusted the model magnitudes of actual saturated images to compensate for the deficiencies noted from 
our simulations.  These adjustments ranged from 0.01~mag for images with one saturated pixel to 0.10~mag for images saturated within
one Airy radius.  The uncertainties of the magnitudes of overexposed stars estimated in this manner are $\sim 0.02$--0.25~mag.  

The locations of sources in the reduced images were determined from either the centroids of the photometric apertures or the pixel
coordinates of the fitted model PSFs.  We estimate that the uncertainties of the former method are 0.1~pixel per axis, and those of
the latter method are 0.2~pixel per axis for the undersampled F110W images and 0.1~pixel per axis for the other filter images.  
(We double these uncertainties for images affected by saturation or bad pixels.)  To determine the angular separations and position 
angles of the binary dwarfs, we adopted the $x$-axis and $y$-axis pixel scales for NIC2 derived by the Space Telescope Science 
Institute around the times of our observations.\footnote{The pixel scales for each NICMOS camera are presently available on the 
World Wide Web at http://www.stsci.edu/hst/nicmos/performance/platescale.}  We estimate that the nominal $1''$ uncertainty in the 
coordinates of {\it HST} guide stars contributes a negliglible error of $\sim 0$\farcs6 to our computed position angles.

\begin{figure}[t]
   \epsscale{1}\plotone{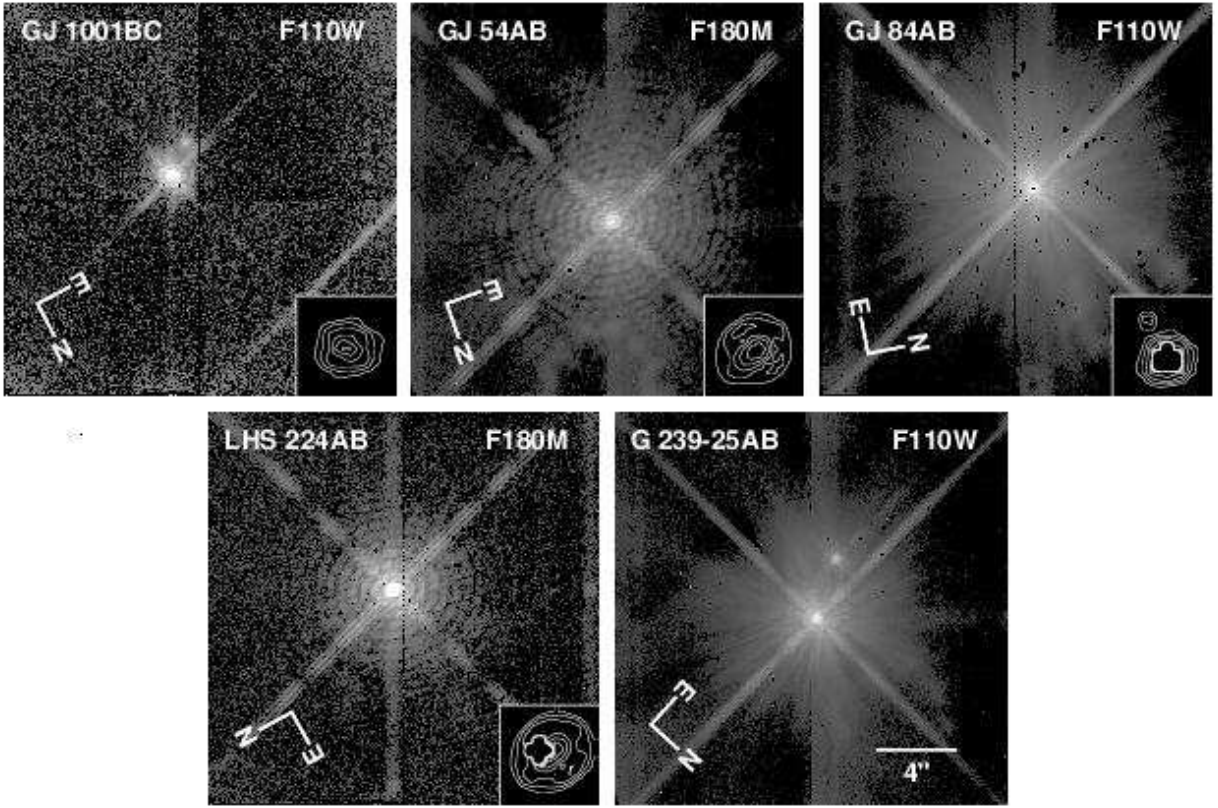}
   \caption{
       NIC2 images of the five dwarfs to which we have detected new or suspected companions.  Each panel contains the 
       filter image that most clearly shows the binary components.  The full NIC2 FOV is displayed with logarithmic pixel 
       scaling, except where masked by labels and insets.  The insets show magnified contour plots of the $13 \times 13$-pixel
       regions of each image centered on the close-binary dwarfs.  Elongated contours indicate the orientations of the marginally 
       resolved components.  The images of GJ~84A and LHS~224AB are affected by overexposure and bad NIC2 pixels, respectively.  
       The faint object located 1\farcs74 to the southeast of GJ~1001BC is an extended source, presumably a background galaxy.
   } 
   \label{Golimowski.fig4}
\end{figure}

Figure~\ref{Golimowski.fig4} shows representative NIC2 images of the five dwarfs to which we have detected new or suspected companions.  The 
insets show magnified contour plots of the four systems separated by less than 0\farcs5.  GJ~1001BC, GJ~54AB, and LHS~224AB are only 
marginally resolved.  The images of LHS~224AB in all four bands are affected by bad NIC2 pixels, and the F110W images of GJ~54AB and 
GJ~84A are saturated.  The faint object located 1\farcs74 to the southeast of GJ~1001BC is an extended source, presumably a background
galaxy.  Table~4 lists the NICMOS magnitudes and relative positions of the components of the five systems, which we determined in the 
manners previously described.  The magnitudes and colors of these dwarfs are depicted as filled circles in Figures~\ref{Golimowski.fig1}a and 
\ref{Golimowski.fig1}b so that they may be readily compared with the NICMOS photometry of other main-sequence stars and brown dwarfs.

\subsection{Follow-up Observations}

Our NICMOS snapshot program is designed to be a single-epoch, multiband survey of all stars within 10~pc.  Determining whether 
these stars and any candidate companions have common proper motion (CPM) technically exceeds the scope of our program.  Nevertheless,
second-epoch observations of candidate companions may be planned at the expense of unobserved targets.  Only 50\% of snapshot
observations are likely to be executed during an {\it HST} observing cycle, however, so these second-epoch observations are not 
guaranteed.  Consequently, we sought more reliable means of confirming our candidates.  We have confirmed four of the five candidate
companions identified in this paper using two telescopes and three instruments, none of which are NICMOS.  We now describe these
confirmatory observations in chronological order.

\subsubsection{GJ~54AB and LHS~224AB}

The apparent separations and brightnesses of GJ~54AB and LHS~224AB (Table~4) make them ideal targets for astrometric monitoring with 
{\it HST's} FGS, which features white-light shearing interferometers capable of measuring the positions of close-binary systems with 
$\sim$~0\farcs001 precision \citep{nel03}.  GJ~54AB was observed with FGS1r in transfer (TRANS) mode on 2000 September~24 and 2003 
June~13.  The TRANS scans were recorded through the wide-band F583W filter.  LHS~224AB was observed with the same instrument configuration
on 2000 January~7 and 2003 October~30.  The first FGS observation of LHS~224AB, which predates our NICMOS observation by three years, was
conducted as part of an independent FGS search for multiplicity among nearby M dwarfs and white dwarfs ({\it HST} Program 8532).  Both 
binary systems continue to be monitored with FGS as parts of {\it HST} Programs 9408, 9972, and 10104.  The complete results of these 
monitoring programs will be presented in subsequent papers.  In the meantime, we present two-epoch results that support the duplicity
of GJ~54 and LHS~224 inferred from our NIC2 images.

\begin{figure}[t]
   \epsscale{1}\plotone{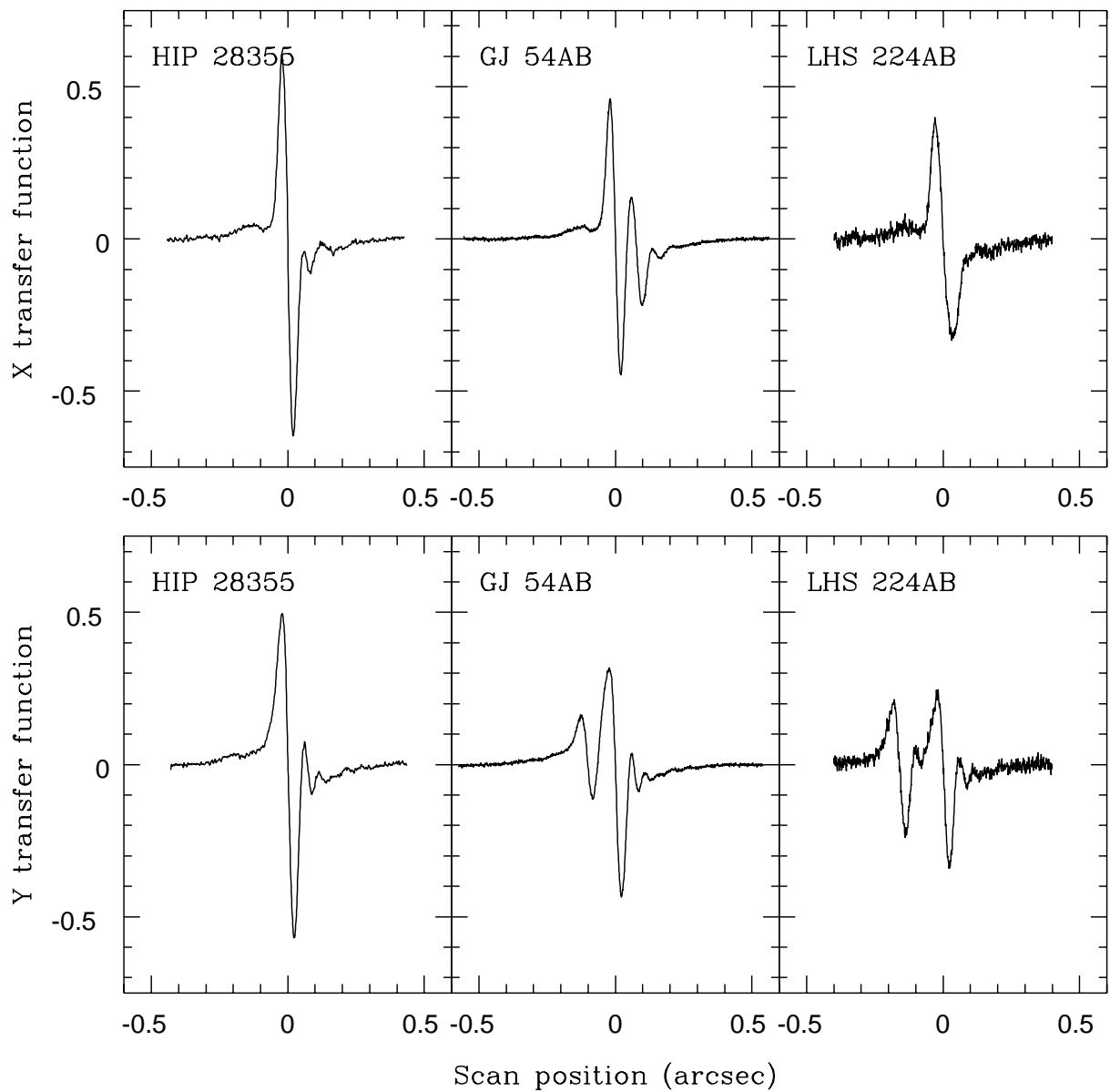}
   \caption{
       Broadband (F583W) FGS1r transfer functions of the single star HIP~28355 {\it (left panels)} and the binary stars GJ~54AB
       {\it (middle panels)} and LHS~224AB {\it (right panels)}.  The observations of GJ~54AB and LHS~224AB were obtained on 2003 
       June~13 and 2003 October~30, respectively.  The transfer functions of single stars are generally antisymmetric about the 
       zero scan positions with peak-to-peak amplitudes of $\sim 1.2$; asymmetric and/or diminished transfer functions reflect the 
       presence of a secondary component.  For instance, the right panels show that LHS~224A and B differ in brightness by 0.3~mag and 
       are separated by 0\farcs03 and 0\farcs16 along the $X$ and $Y$ axes, respectively.
   }
   \label{Golimowski.fig5}
\end{figure}

The TRANS-mode observations of each binary star were calibrated and processed in the manner described by \citet{fra91}.  Figure~\ref{Golimowski.fig5}
shows the transfer functions of GJ~54AB, LHS~224AB, and the single star HIP~28355, measured along orthogonal axes of FGS1r.  These $X$ and 
$Y$ transfer functions track the visibility of the broadband fringes as the stars are scanned across the Koester's prism \citep{fra92}.  
The transfer functions of a single star (left panels of Figure~\ref{Golimowski.fig5}) are generally antisymmetric about the zero scan position that 
denotes alignment between the incoming wavefront and the optical axis.  The transfer function of a binary star along a given FGS axis is 
the superposition of the individual stars' transfer functions, offset by their projected separation along that axis.  The transfer 
functions of LHS~224AB recorded on 2003 October~30 (right panels of Figure~\ref{Golimowski.fig5}) show that the binary star's orientation was nearly 
parallel to the $Y$-axis of FGS1r and its separation was $\sim 0\farcs16$.  Likewise, the middle panels of Figure~\ref{Golimowski.fig5} show that, 
on 2003 June~13, GJ~54A and B were separated by $\sim 0\farcs13$ and were oriented obliquely to FGS1r's $X$ and $Y$ axes.

Decomposing the $X$ and $Y$ transfer functions of GJ~54AB and LHS~224AB into their constituent transfer functions allows us to determine 
the relative positions and F583W magnitudes of their components \citep{fra91}.  Table~5 lists our preliminary results.  The differences 
between the F583W magnitudes can be transformed to Johnson $V$-band magnitude differences ($\Delta V$) in the manner described
by \citet{hen99}.  Employing this technique, we estimate $\Delta V = 1.03 \pm 0.08$ and $0.29 \pm 0.02$ for GJ~54AB and LHS~224AB, 
respectively.  These values reflect the averages of the measurements obtained along each FGS axis at two epochs of observation.  Their 
errors include those associated with the deconvolution of the transfer functions, the photometric transformations, and the published 
parallaxes and composite $V$ magnitudes of each system.

\subsubsection{G~239--25AB}

Because the components of G~239--25 are relatively bright and widely separated (Table~4), follow-up observations of the system were 
possible using conventional ground-based imaging.  We observed G~239--25AB on 2003 May~06 using the ARC 3.5~m telescope and the Seaver 
Prototype Imaging Camera (SPICam), which features a $2048 \times 2048$-pixel CCD with an unbinned image scale of 0\farcs141~pix$^{-1}$.
Exposures of 1~s, 10~s, and 60~s were recorded using the Sloan Digital Sky Survey (SDSS) $z$ filter and $2 \times 2$ pixel binning.  
The 1~s image of G~239--25A was unsaturated and indicated a seeing-limited image resolution of 1\farcs25.  The sky transparency was 
variable, so no photometric calibration was possible.  The images were reduced conventionally by subtracting the electronic bias and 
dividing by a flat-field image.

\begin{figure}[t]
   \epsscale{1}\plotone{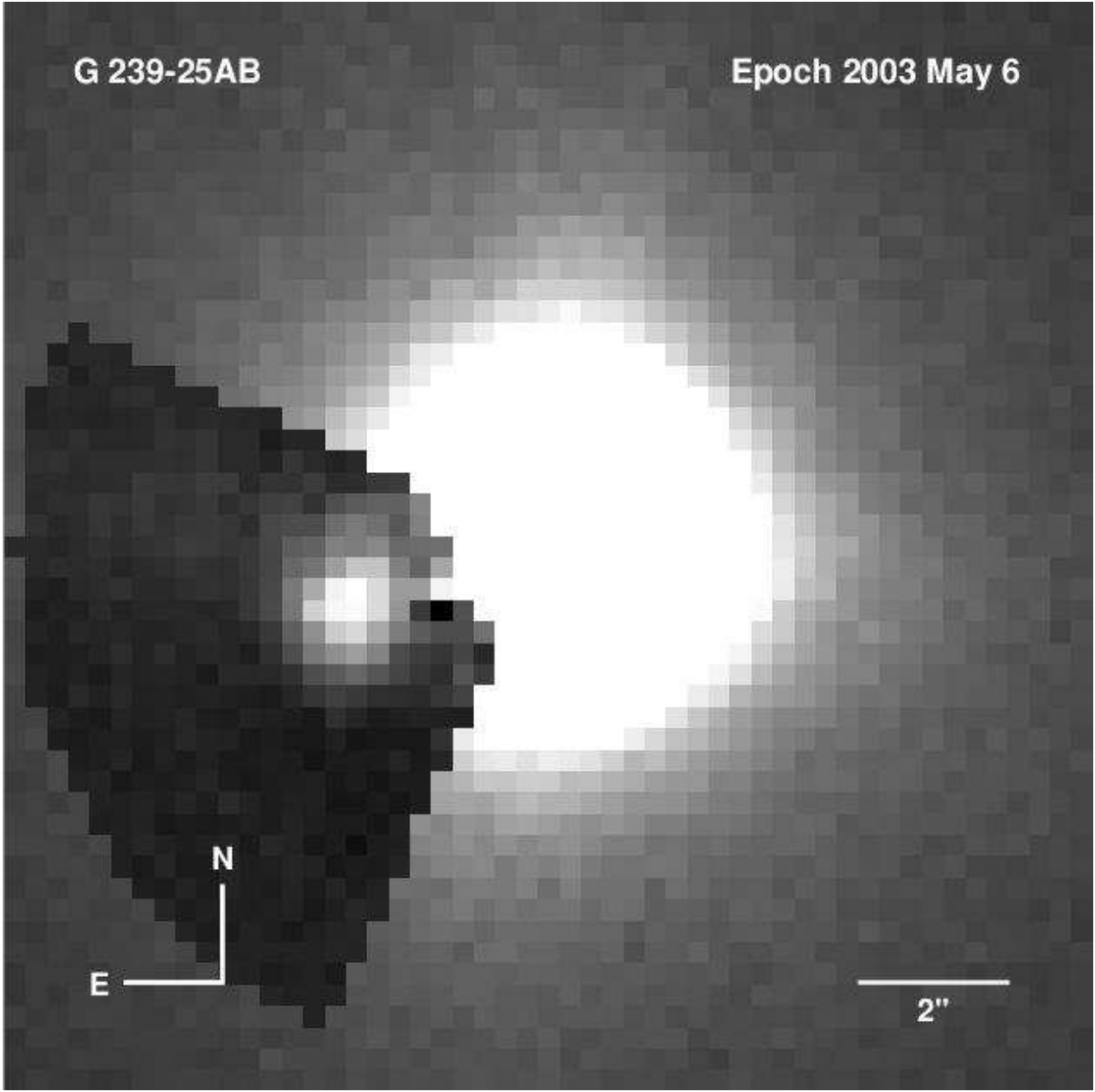}
   \caption{
       Image of G~239--25AB recorded through the SDSS $z$ filter on 2003 May~06 using the ARC 3.5~m telescope and SPICam at Apache 
       Point Observatory.  This $14\farcs4 \times 14\farcs4$ region of the 1~s image is displayed with a linear scale and clipped
       at $1000~e^-$ to limit the dynamic range.  The seeing-limited PSF of \hbox{G~239--25A} has been partly subtracted to reveal
       more clearly the faint companion located 2\farcs84 to the east--southeast.
   }
   \label{Golimowski.fig6}
\end{figure}

Figure~\ref{Golimowski.fig6} shows a $14\farcs4 \times 14\farcs4$ section of the the reduced 1~s image centered on G239--25A.  Part of the 
star's seeing-limited PSF has been subtracted to reveal G239--25B more clearly.  [Lacking a comparably exposed reference star, we 
subtracted artificial background values that were interpolated from radial profiles of the PSF at azimuths $\pm 45^{\circ}$ from 
the position angle (PA) of G239--25B.]  The companion is located $2\farcs84 \pm 0.14$ from G~239--25A at PA~$= 106^{\circ}\!{.}5 
\pm 3^{\circ}\!{.}6$.  The errors in these values include a 0.5~pixel uncertainty in each coordinate of G~239--25B's centroid and a 
possible $0^{\circ}\!{.}4$ error in the alignment of SPICam's detector with the celestial axes (R.~McMillan 2004, private communication).
The integrated signals of each star within circular apertures of radius 0\farcs85 (3 pixels) indicate a magnitude difference of 
$\Delta z \approx 4.9$.

\subsubsection{GJ~1001ABC}

A follow-up observation of the GJ~1001 triple system was performed on 2003 August~20 using the High Resolution Channel (HRC) of 
the {\it HST} ACS \citep{for03,pav03}.\footnote{This observation was conducted as part of the guaranteed observing time awarded to 
the ACS Investigation Definition Team.}  The HRC features a $1024 \times 1024$-pixel CCD with a pixel scale of $\sim 0$\farcs$025 
\times 0$\farcs$028$.  Its $\sim 26'' \times 29''$ FOV allowed simultaneous imaging of all three components of the system.  Two 
sets of 2~s, 60~s, and 300~s exposures were recorded through the F850LP (SDSS $z$) filter to allow unsaturated imaging of each 
component.  Between exposures, the telescope was offset by $\sim 6.5$ pixels along each CCD axis to diminish the possibility of 
image corruption by bad pixels.  Likewise, two sets of 2~s and 300~s exposures were recorded through the F625W (SDSS $r$) and 
F775W (SDSS $i$) filters.  All exposures were recorded using an analog-to-digital conversion gain of $2.2~e^-$~DN$^{-1}$.  The 
digitized images were bias-subtracted and flattened using the standard ACS calibration software \citep{mac03} and appropriate 
reference images and tables.  The flattened images were then processed with the ACS Pipeline Science Investigation Software 
\citep[APSIS;][]{bla03} to remove cosmic-ray contaminants, correct the images for geometric distortion, and combine the dithered 
pairs of images.  The reduced and resampled HRC images have rectified pixel scales of 0\farcs025 along each axis.  

\begin{figure}[t]
   \epsscale{1}\plotone{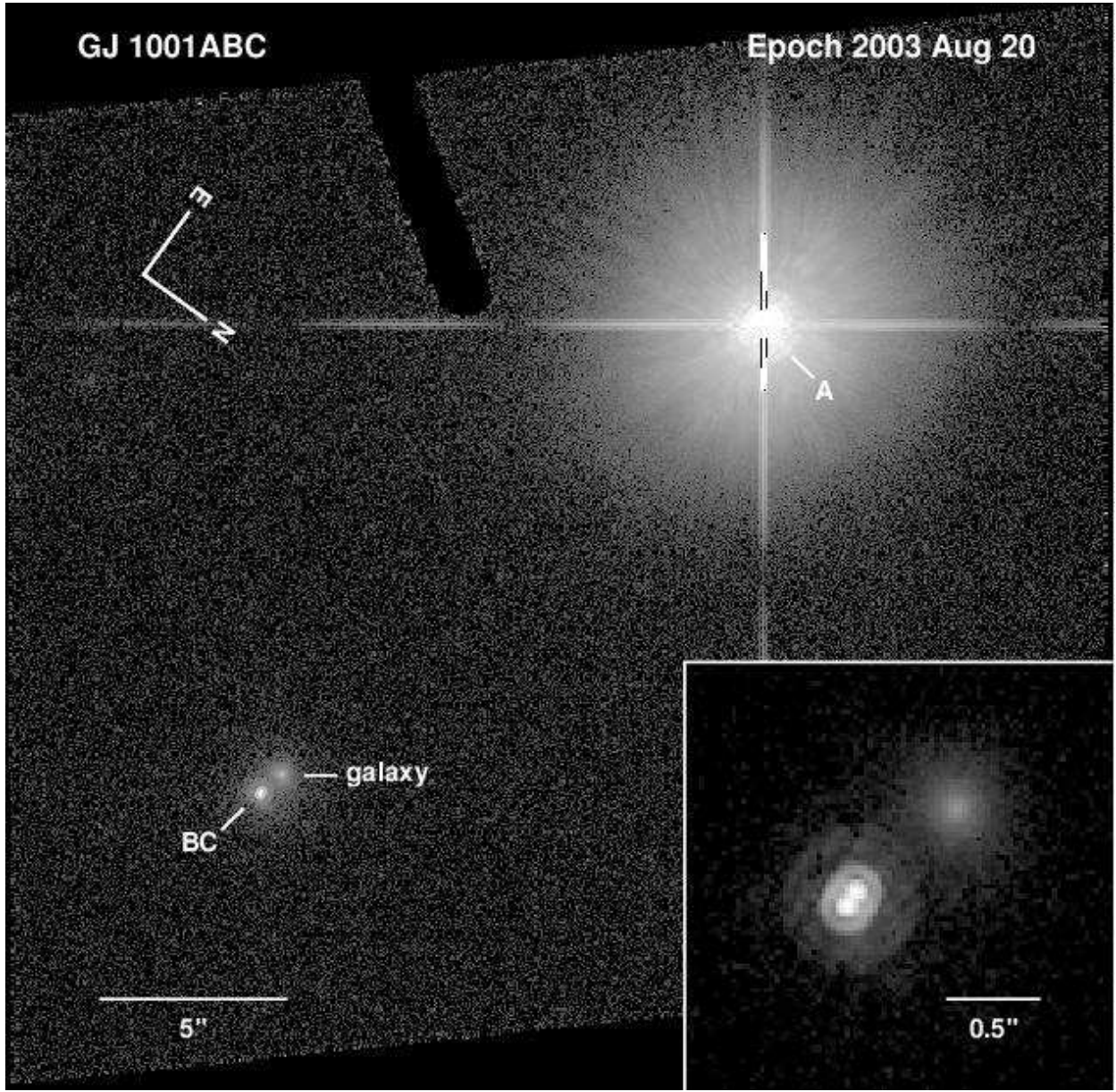}
   \caption{
       F775W (SDSS $i$) image of the GJ~1001 triple system obtained with the ACS HRC and processed by APSIS (see \S2.2.3).  
       The rhomboidal image is a geometrically corrected composite of two dithered 300~s HRC exposures.  The binary L dwarf 
       GJ~1001BC lies $\sim 18\farcs2$ from the M3.5 primary star GJ~1001A and $\sim 0\farcs75$ from the same anonymous galaxy
       seen in the NIC2 image (Figure~\ref{Golimowski.fig4}).  No other sources appear in this high-latitude ($b = -73^{\circ}$) field.  
       The inset shows a magnified and rescaled region centered on GJ~1001BC and the galaxy.   The position angle of 
       GJ~1001BC has changed significantly since the NIC2 observation of 2002~October~11.
   }
   \label{Golimowski.fig7}
\end{figure}

Figure~\ref{Golimowski.fig7} shows the resultant $2 \times 300$~s F775W image of GJ~1001ABC.  Although GJ~1001A is overexposed in this image,
it is suitably exposed in our 2~s F775W images.  These levels of exposure are also characteristic of the images obtained with the 
F625W and F850LP filters.  The binary L dwarf is clearly resolved, and its photocenter lies $\sim 18$\farcs2 from GJ~1001A at 
PA~$\approx 259^{\circ}$.  These values are consistent with the relative positions of GJ~1001A and B reported by \citet{gol99} from 
lower resolution, ground-based images in which GJ~1001A is overexposed.  

Although the angular resolution of the HRC at $\lambda = 0.9~\mu$m is $\sim 2.2$ times better than that of NIC2 at $\lambda = 
1.1~\mu$m, the HRC images of GJ~1001B and C overlap sufficiently to preclude the use of aperture photometry.  Consequently, we 
generated model PSFs for each component using Tiny Tim \citep{kri03} and a reference spectrum of the L4 dwarf 2MASS~J03454316+2540233
\citep{geb02}.  Because Tiny Tim produces only geometrically distorted PSFs, we fitted model PSFs to the flattened but distorted 
(i.e., pre-APSIS) images of GJ~1001B and C.  Tiny Tim does not fully simulate the diffuse halo of scattered light that is superposed 
on point-source images longward of $\sim 0.8~\mu$m \citep{sir98}, so we derived approximate corrections for this effect in the F775W
and F850LP bandpasses for each dwarf.  In doing so, we normalized the integrated pixel values from the PSFs fitted to the images of 
GJ~1001A with those obtained directly from the APSIS-processed images using circular apertures with radii of 0\farcs5 and the 
aperture corrections for M dwarfs reported by \citet{sir04}.  We then applied the same normalization to the PSFs fitted to the 
images of GJ~1001B and C along with additional corrections of $\Delta m_{\rm F775W} = -0.008$ and $\Delta m_{\rm F850LP} = -0.057$ 
to compensate for the spectral differences between GJ~1001A and GJ~1001BC \citep{sir04}.  We also applied small ($\sim 1$\%) 
corrections for photometric losses caused by CCD charge-transfer inefficiency \citep{rie03}.  We estimate that the uncertainties of 
these corrected count rates are $\sim 4$--8\%.  We converted the count rates to AB and Vega-based magnitudes using the photometric 
calibration recipe of \citet{sir04}, which specifies that Vega has zero magnitude in all bandpasses.  Table~6 lists the resulting 
magnitudes and errors of GJ~1001ABC in each bandpass.

To determine the relative separations and orientations of the system components, we transformed the pixel coordinates of the fitted
PSFs (which are measured from the geometrically distorted images) to angular coordinates measured along the V2 and V3 axes that
define {\it HST's} principal focal plane.  In doing so, we used the latest image-distortion correction table (IDCTAB) for the HRC, 
which has an RMS residual error of 0.03--0.05~pixels per axis for the filters used in our study \citep{meu02}.  The errors in the 
measured centroids of our fitted PSFs are $\sim 0.1$~pixel, and the uncertainty in the orientation of the HRC FOV due to the $\pm 1''$
inaccuracy of {\it HST} guide star coordinates is $\pm 0$\farcs7.  Columns 6--8 of Table~6 list the apparent and projected separations
and position angles of the GJ~1001 system, based on the average measurements from individual HRC images and the parallax given in 
Table~1.  The first set of values pertains to the separation and orientation of the photocenter of GJ~1001BC with respect to GJ~1001A.
The second set of values pertains to components B and C separately.

\section{Discussion}

The PSF-subtracted NIC2 images of each dwarf reveal no sources other than the five companions and the single galaxy in the field of 
GJ~1001BC.  The paucity of background sources is not suprising given the generally high Galactic latitudes of the dwarfs (col.~4 of 
Table~1) and NIC2's small FOV.  Thus, the dwarfs' high latitudes raise the likelihood that any apparent companions are true companions.
Indeed, the high latitudes of these dwarfs prompted their inclusion in this paper announcing the first confirmed companions of 
our NICMOS survey.  We now discuss the evidence confirming these companions and the implications for our future understanding of VLM 
multiple systems.  We begin with GJ~1001BC, which is arguably the most significant of the newfound binaries from the perspective of 
constraining the MLR in the substellar regime.

\subsection{GJ~1001ABC}

Assuming that GJ~1001C is always the fainter of the close-binary components seen in our NIC2 and HRC images, then GJ~1001C moved 
$\sim 0\farcs07$ at PA~$\sim 164^{\circ}$ relative to GJ~1001B during the 313 days between our epochs of observation.  This 
motion is much less than that expected from the annual proper motion of the GJ~1001 system (1\farcs618 at PA~$= 154^{\circ}\!{.}5$; 
\citealt{van95}).  However, the NIC2 magnitudes of GJ~1001B and C differ by amounts similar to their combined uncertainties, so their
identities at each epoch are somewhat ambiguous.  If we switch their identities in the NIC2 images, then GJ~1001C moved $\sim 0\farcs16$
at PA~$\sim 72^{\circ}$ relative to GJ~1001B between our NICMOS and ACS observations.  This motion is also much smaller and differently
oriented than the annual proper motion of the GJ~1001 system.  Conversely, the photocenter of GJ~1001BC moved 1\farcs41 at PA~$= 
153^{\circ}\!{.}6$ relative to the galaxy labeled in Figure~\ref{Golimowski.fig7}, which is consistent with the annual proper motion.  We 
therefore conclude that GJ~1001A, B, and C have CPM regardless of how the components are labeled, and that the relative motion of 
GJ~1001B and C is orbital in nature.

Using the observed separations of GJ~1001BC, a distance to the triple system of $d = 9.55 \pm 1.04$~pc \citep{van95}, and the statistical 
relation between the observed angular separation ($\alpha$) and the semimajor axis of the relative orbit ($a_{\rm rel}$),

\begin{equation} 
\langle a_{\rm rel} \rangle = 1.26~d~\langle\alpha\rangle
\end{equation}

\noindent
\citep{fis92}, we estimate that $a_{\rm rel} \approx 1.04 \pm 0.12$~AU.  The age of GJ~1001A is loosely constrained to 1--10~Gyr \citep{leg02b,
giz02}, so we infer from evolutionary models \citep{bur97,cha00} that its coeval L4.5 dwarf companions \citep[$T_{\rm eff} \approx 1750$~K;]
[]{gol04} have masses of 0.06--0.075~$M_{\odot}$.  If so, then the period of GJ~1001BC's orbit is $\sim 2.9 \pm 0.5$~yr, which is the 
shortest estimated period of any known binary L or T dwarf (Table~7).  However, the $\sim 47^{\circ}$ (clockwise) difference between the 
position angles of GJ~1001C in the NIC2 and HRC images suggests either that the orbital period is actually much longer than 3~yr, or that 
GJ~1001B and C are not consistently identified in the NIC2 and HRC images.  If the identities of GJ~1001B and C in the NIC2 images are 
switched, then GJ~1001C's position angle changed by $\sim 133^{\circ}$ (anticlockwise), which is more compatible with a $\sim 3$~yr period
than the motion inferred from our adopted designations of GJ~1001B and C.

Since its discovery in 1999, GJ~1001B has been well studied photometrically and spectroscopically \citep{gol99,bas00,leg02b}.  As a 
companion to an M dwarf of known distance, it has frequently been used to study the relationships between the luminosities, colors, and
spectral types of L dwarfs \citep{kir00,dah02,kna04,gol04}.  None of these studies describes GJ~1001B as overluminous for its color
or spectral type, as would be expected for a non-eclipsed binary system with components of equal luminosity.  The fact that GJ~1001BC
is binary indicates that either its components are underluminous or its inferred trigonometric parallax is incorrect.  The 
composite optical and near-infrared spectra of GJ~1001BC show that the components are typical mid-L dwarfs and reveal no cause for 
underluminosity in any optical or near-infrared bandpass.  Applying the SDSS color--magnitude relation of \citet{haw02} to our ACS 
photometry of each component (Table~6), we compute an average photometric parallax of $0\farcs046 \pm 0\farcs017$, or a distance 
to GJ~1001BC of $21.8 \pm 7.9$~pc.  Similarly applying the spectroscopic parallax relations of \citet{haw02} for L4.5 dwarfs, we
obtain distance moduli of $M$--$m \approx -0.4$ to $-0.7$, or $d \approx 12$--14~pc.  Although these estimated photometric and 
spectroscopic distances agree only marginally, both suggest that the only published trigonometric parallax of GJ~1001A \citep{van95}
is too large.  Indeed, preliminary astrometric results for GJ~1001A from the Cerro Tololo Parallax Investigation indicate that the 
star's distance is closer to 15~pc than to 10~pc (T.~Henry et al., in preparation).

Assuming a revised distance of $\sim 15$~pc to GJ~1001BC, we estimate that its relative orbit has a semimajor axis of $\sim 1.6$~AU 
and a period of $\sim 5.5$~yr.  Unfortunately, the uncertainties associated with these estimates prevent us from identifying GJ~1001B
and C in the NIC2 images unambiguously.  Regardless of the components' designations, GJ~1001BC offers a superb opportunity to 
determine the orbit and dynamical masses of two probable brown dwarfs in a small amount of time.  Our HRC images show that {\it HST} 
can provide the high-resolution images necessary for such pioneering measurements, if its lifetime is extended by a fifth servicing 
mission.  GJ~1001BC's faintness and small separation pose a challenge for ground-based telescopes equipped with adaptive optics (AO),
but the image resolution achieved during recent AO observations of the binary T dwarfs $\epsilon$~Indi~BC \citep{mcc04} indicate that 
long-term monitoring of GJ~1001BC's orbit should be possible with the latest ground-based imaging technology.

\subsection{GJ~54AB}

The suspected duplicity of GJ~54 was first reported by \citet{rod74}, who visually observed a ``very close double star of nearly 
equal components in the 40-inch reflector'' at Siding Spring Observatory.  They provided no information about the separation of 
the double star or the epoch of their observation, but the lack of other stars of comparable brightness within several arcminutes
of GJ~54 suggests that \citeauthor{rod74} did observe the same pair of stars seen in our NIC2 images.  The $V$-band diffraction limit
of a 40-inch telescope is comparable to the separation of GJ~54AB measured from our NIC2 and FGS data, so \citeauthor{rod74} may have 
seen the pair during instances of excellent atmospheric seeing.  Therefore, we credit \citet{rod74} with the initial discovery of 
GJ~54AB, and reserve for ourselves the pleasure of confirming and quantifying that discovery.

The FGS TRANS scans of GJ~54 recorded on 2000 September~24 and 2003 June~13 (Figure~\ref{Golimowski.fig5}) show two components separated by $\sim
0\farcs12$ (Table~5).  This separation is similar to that measured from our NIC2 images obtained on 1998 November~9, and it is much less than
GJ~54's annual proper motion of $0\farcs692$ \citep{esa97}.  We therefore conclude that GJ~54A and B exhibit CPM and are physically bound.
Figures~\ref{Golimowski.fig1}a and \ref{Golimowski.fig1}b show that the NIC2 magnitudes and colors of GJ~54A and B are consistent with other stars in our 
10~pc sample having spectral types M2.0--M2.5 and M3.0, respectively.  Using our computed value of $\Delta V$ for GJ~54AB (\S2.2.1) and 
published measurements of the stars' composite $V$ magnitude and parallax (Table~1), we obtain $M_V = 10.60 \pm 0.11$ and $M_V = 11.63 
\pm 0.12$ for GJ~54A and B, respectively.  Employing the $V$-band MLR of \citet{del00}, we compute masses of $0.42 \pm 0.07~M_{\odot}$ 
and $0.31 \pm 0.07~M_{\odot}$ for the two stars.

Using our {\it HST} measurements, a distance to GJ~54AB of $8.20 \pm 0.41$~pc (Table~1), and equation~(2), we estimate a semimajor axis of 
$1.33 \pm 0.09$~AU and a period of $1.8 \pm 0.2$~yr for GJ~54AB's relative orbit.  After deconvolving the individual components of the 
FGS transfer functions and transforming them to celestial positions, we find that the position of GJ~54B relative to GJ~54A on 2000 
September~24 was nearly coincident with the position observed in the NIC2 images recorded 1.87~yr before.  In fact, a preliminary fit of 
a Keplerian orbit to our three-epoch astrometric measurements yields a highly inclined orbit with a semimajor axis of $\sim 0\farcs17$
($\sim 1.4$~AU), a period of $\sim 1.75$~yr, and masses of $0.49~M_{\odot}$ and $0.33~M_{\odot}$.  Thus, our crude photometric and 
astrometric estimates of the orbits and masses of GJ~54AB are mutually consistent.

\subsection{GJ~84AB}

\citet{nid02} reported 11 radial-velocity measurements of GJ~84 obtained at the 10~m Keck~I Observatory from 1998 February to 2001 July 
that indicate the presence of a low-mass companion.  They reproduced the velocity variations by fitting a partial Keplerian orbit with a 
period of $19 \pm 7$~yr, a semiamplitude of $2.2 \pm 0.2$~km~s$^{-1}$, and an eccentricity of $0.44 \pm 0.08$.  They also estimated for 
the companion a minimum mass of $0.115~M_{\odot}$ and a minimum semimajor axis of 5.6~AU.  These values are consistent with the luminosity
and projected separation of the faint companion seen in our NIC2 images (Table~4).  Our images thus confirm the duplicity of GJ~84, and 
are useful for further constraining the parameters of the binary orbit.

Figure~\ref{Golimowski.fig1}b shows that the values of $M_{\rm F110W} = 11.10 \pm 0.10$ and F110W--F222M~$= 1.58 \pm 0.10$ measured for GJ~84B lie 
between those of the M7 dwarf GJ~105C \citep[$M_{\rm F110W} = 10.93$, F110W--F222M~$= 1.42$; revised from][]{gol00} and the M8 dwarf GJ~752B 
($M_{\rm F110W} = 11.59$, F110W--F222M$~= 1.71$).  The anomalous location of GJ~84B's colors in Figure~\ref{Golimowski.fig1}a is likely due to a 
singularly large error in our F180M measurement caused by the coincidence of GJ~84B's image and a diffraction spike from GJ~84A 
(Figures~\ref{Golimowski.fig2} and \ref{Golimowski.fig3}).  As the magnitudes and colors of GJ~84B generally match those of 
GJ~105C within their combined uncertainties, we tentatively assign a spectral type of M7 to GJ~84B.   To estimate the masses of GJ~84A and B
from current near-infrared MLRs, we loosely equate our measured values of $M_{\rm F110W}$ and $\frac{1}{2} (M_{\rm F207M} + M_{\rm F222M})$ 
for each star with its conventional $J$- and $K$-band absolute magnitude, respectively.  Employing the MLRs of \citet{del00}, we obtain masses
of $0.51 \pm 0.06~M_{\odot}$ and $0.084 \pm 0.026~M_{\odot}$ for GJ~84A and B, respectively.  Using these masses, a distance to GJ~84AB of 
$9.40 \pm 0.17$~pc (Table~1), the observed binary separation (Table~4), and equation~(2), we estimate a semimajor axis of $5.25 \pm 0.12$~AU
and a period of $15.6 \pm 1.0$~yr for GJ~84AB's relative orbit.

We can investigate the consistency between the orbital characteristics estimated from our NIC2 images and those derived by \citet{nid02}
from the Keplerian fit to their measured radial velocities.  Kepler's Third Law may be expressed as 

\begin{equation}
\frac{a_1}{\pi_{\rm abs}} = P^{2/3} \frac{M_2}{(M_1 + M_2)^{2/3}}
\end{equation}

\noindent
\citep{lan80}, where $a_1$ is the semimajor axis of the primary star's orbit about the barycenter in units of arcsec, $\pi_{\rm abs}$ is the 
absolute parallax, $P$ is the orbital period expressed in years, and $M_1$ and $M_2$ are the primary and secondary masses in units of 
$M_{\odot}$.  Using our estimated masses of GJ~84A and B and the fitted radial-velocity period of \citet{nid02}, we obtain $a_1 \approx 
0\farcs089 \pm 0\farcs035$.  An equation relating orbital parameters derived astrometrically and spectroscopically \citep{lan80} can be 
expressed as

\begin{equation}
\frac{a_1~{\rm sin}~i}{\pi_{\rm abs}} = \frac{P K_1 \sqrt{1 - e^2}}{2 \pi \times 4.7405},
\end{equation}

\noindent
where $i$ is the inclination of the orbit with respect to the line of sight, $K_1$ is the semiamplitude of the primary star's radial-velocity 
orbit in units of km~s$^{-1}$, $e$ is the eccentricity of the orbit, and $\pi \approx 3.14159$.  Inserting the values of $a_1$, $P$, $K_1$, 
and $e$ derived by \citet{nid02} and us, we obtain ${\rm sin}~i \approx 1.49 \pm 0.85$.  Only the values of ${\rm sin}~i$ deviating by more than 
$-0.5 \sigma$ from 1.49 are physical, which indicates that the Keplerian fit to the radial velocity data is only marginally compatible with our 
observed luminosities (masses) of GJ~84AB.  Clearly, more spectroscopic and astrometric observations are needed over the next decade to better 
constrain the orbit of GJ~84AB.

\subsection{LHS~224AB}

The duplicity of LHS~224 was first suspected by \citet{dah82}, who noted a possible astrometric perturbation with a period of $\sim 6$~yr
from photographic plates produced by the United States Naval Observatory from 1969 to 1980.  After another decade of astrometric monitoring, 
\citet{har93} could neither confirm nor deny this suspicion.  In 1999, LHS~224 was included in our ongoing FGS survey of nearby and possibly 
binary red and white dwarfs.  The TRANS scans of LHS~224 recorded on 2000 January~7 and 2003 October~30 (Figure~\ref{Golimowski.fig5}) show two 
components separated by $\sim 0\farcs16$ (Table~5).  This separation is similar to that measured from our NIC2 images obtained on 2003 
March~13, and it is much less than LHS~224's annual proper motion of $1\farcs166$ \citep{gli91}.  We therefore conclude that LHS~224A and B 
exhibit CPM and are physically bound.  

Figures~\ref{Golimowski.fig1}a and \ref{Golimowski.fig1}b suggest that the NIC2 F110W magnitudes and F110W--F222M colors of LHS~224AB are brighter and bluer, 
respectively, than those of other M5 dwarfs.  Such anomalous behavior is not seen in other color--magnitude and color--color diagrams 
constructed from our F180M, F207M, and F222M photometry of nearby stars.  We therefore suspect that our F110W measurements are adversely 
affected by the undersampled F110W images of LHS~224AB and by the unfortunate proximity of bad pixels to those images (Figure~\ref{Golimowski.fig4}).
The magnitudes and colors of LHS~224A and B measured in the other bandpasses are consistent with those of several nearby dwarfs having types 
M4.5--M5.0 and M5.0--M5.5, respectively.  Using our computed value of $\Delta V$ for LHS~224AB (\S2.2.1) and published measurements of the 
stars' composite $V$ magnitude and parallax (Table~1), we obtain $M_V = 14.08 \pm 0.05$ and $M_V = 14.37 \pm 0.05$ for LHS~224A and B,
respectively.  Employing the $V$-band MLR of \citet{del00}, we compute masses of $0.16 \pm 0.02~M_{\odot}$ and $0.14 \pm 0.02~M_{\odot}$ for
the two stars.

Using our NIC2 measurements, a distance to LHS~224AB of $9.2 \pm 0.2$~pc \citep{van95}, and equation~(2), we estimate $a_{\rm rel} \approx 
1.53 \pm 0.08$~AU and $P \approx 3.5 \pm 0.3$~yr for LHS~224AB's orbit.  A preliminary fit of a Keplerian orbit to our three-epoch
astrometric measurements yields $a_{\rm rel} \approx 1.6$~AU, $P \approx 3.3$~yr, and masses of $\sim 0.21~M_{\odot}$ and $\sim 0.20~M_{\odot}$.
These dynamic masses exceed our photometric estimates by $3 \sigma$, but we expect these discrepancies to diminish greatly as more 
FGS measurements are obtained over a broader range of orbital phases.

\subsection{G~239--25AB}

During the 4.5 years between our NIC2 and SPICam observations, G~239--25A's proper motion was 1\farcs42 toward PA~$= 259^{\circ}\!{.}2$ 
\citep{esa97}.  The magnitude and direction of this motion are inconsistent with the small differences between the relative positions of 
G~239--25A and B at each epoch.  We therefore conclude that G~239--25A and B have CPM and are gravitationally bound.  Moreover, we believe
that the observed motion of G~239--25B relative to G~239--25A ($0\farcs43 \pm 0\farcs14$ toward PA~$= 352^{\circ} \pm 22^{\circ}$) is 
orbital in nature.

Figures~\ref{Golimowski.fig1}a and \ref{Golimowski.fig1}b show that the NIC2 magnitudes and colors of G~239--25B are similar to those of the M8 dwarf GJ~752B
and the spectroscopically unclassified late-M dwarf GJ~1245C.  The latter dwarf is the least massive M dwarf \citep[$0.074 \pm 
0.013~M_{\odot}$;][]{hen99} for which a mass has been computed exclusively from dynamical elements.\footnote{\citet{lan01} mapped the relative 
orbit of the M8.5+M9 dwarf binary GJ~569BC and derived a total mass of $0.123 \pm 0.025~M_{\odot}$ for the two dwarfs.  Lacking a dynamical
measurement of the components' mass ratio, they inferred a value of 0.89 from their relative near-infrared photometry of the pair.}  We 
assume for our present purpose that G~239--25B has the same spectral type and mass as GJ~752B and GJ~1245C, respectively.  Liberally 
applying our F110W, F207M, and F222M photometry of G~239--25A to the near-infrared MLRs of \citeauthor{del00}~(2000; see \S3.3 of this 
paper), we estimate a mass of $0.32 \pm 0.15~M_{\odot}$ for the M3 dwarf.  Using these masses, a distance to G~239--25AB of $9.9 \pm 1.2$~pc
\citep{esa97}, the observed binary separations (Table~4 and \S2.2.2), and equation~(2), we estimate $a_{\rm rel} \approx 37.8 \pm 4.8$~AU 
and $P \approx 370 \pm 100$~yr for G~239--25AB's orbit.

If our estimates of the physical and dynamical characteristics of G~239--25AB are correct, then G~239--25A's radial velocity should vary 
periodically with an amplitude of $\sim 3$~km~s$^{-1}$.  \citet{nid02} reported that the radial velocity of G~239--25A (HIP~71898) was 
stable to within 0.1~km~s$^{-1}$ of 18.704~km~s$^{-1}$ over 864 days distributed about a mean calendar date of 1999 January~5.  On the 
other hand, \citet{upg96} reported that the star's radial velocity increased from 17.51 to 21.17~km~s$^{-1}$ (with an average uncertainty
of $\pm 1.22$~km~s$^{-1}$) over 68 days in mid-1994.  This trend is hard to explain in the face of the more precise measurements of 
\citet{nid02} and our estimated period of $\sim 370$~yr.  The {\it Hipparcos} and Tycho Catalogues note that the large error associated
with the G~239--25A's parallax (Table~1) is consistent with the effects of a short-period astrometric binary \citep{esa97}.  However, 
subsequent reanalysis of the {\it Hipparcos} measurements yielded a ``more likely'' single-star parallax solution of $92.62 \pm 1.52$~mas
\citep{esa97}.  Thus, the results of \citet{upg96} are probably not caused by unresolved duplicity.

\subsection{Implications of New and Confirmed Companions}

The companions described in this paper increase the number of known binary systems within 10~pc by 8\% and the number of known systems 
within 10~pc having three or more components by 6\% (T.~Henry et al., in preparation).  These new and confirmed systems can be grouped 
in several ways: 

\begin{itemize}
\item Orbital periods $\lesssim 5$~yr (GJ~1001BC, GJ~54AB, and LHS~224AB)
\item Component masses $\lesssim 0.15~M_{\odot}$ (GJ~1001BC, GJ~84B, LHS~224AB, and G~239--25B)
\item Secondary/primary mass ratios $\lesssim 0.25$ (GJ~84AB and G~239--25AB)
\item Secondary/primary mass ratios $\gtrsim 0.75$ (GJ~1001BC, GJ~54AB, and LHS~224AB)
\end{itemize}

Astrometrically monitoring the first group will quickly yield dynamic masses needed to constrain the MLR at well-spaced intervals between 
0.50 and $0.05~M_{\odot}$.  Monitoring the second group will extend the presently known MLR beyond its lower limit of $\sim 0.1~M_{\odot}$,
where the effects of varying metallicity and age are pronounced \citep{bar98}.  As members of the first two groups, GJ~1001BC and LHS~224AB 
merit immediate and comprehensive follow-up study.  Indeed, GJ~1001BC offers one of the best opportunities to date for determining empirically
the masses of binary brown dwarfs.\footnote{\citet{bou04} recently determined a total dynamical mass of $0.146^{+0.016}_{-0.006}~M_{\odot}$ 
for the L0+L1.5 binary 2MASS~J07464256+2000321AB from its relative orbit.  Using photometric and astrometric data and theoretical evolutionary 
tracks, they estimated masses of $0.085 \pm 0.010~M_{\odot}$ and $0.066 \pm 0.006~M_{\odot}$ for the L0 and L1.5 dwarfs, respectively.}  
GJ~84AB provides an 
opportunity for determining the mass of an M dwarf near the hydrogen-burning limit, but the system's comparatively long period ($\sim 15$~yr)
requires a greater commitment of resources and patience than is needed for monitoring GJ~1001BC, GJ~54AB, and LHS~224AB.  Unfortunately, an
accurate measurement of the mass of the candidate brown dwarf G~239--25B probably will not be obtained within the lifetimes of current 
generations of astronomers. 

Comparing the masses of GJ~1001BC with those of other short-period binary brown dwarfs having different spectral types (Table~7) will be the 
first steps toward empirically mapping the cooling tracks of brown dwarfs with various masses.   Moreover, by assuming that GJ~1001BC and other
suspected substellar members of close-multiple systems [such as GJ~1245C \citep{hen99}, the M8.5--M9.5 dwarfs LHS~1070BC \citep{lei01} and 
GJ~569BC \citep{mart00,lan01}, the L dwarfs GJ~417BC \citep{bou03} and GJ~564BC \citep{pot02}, and the T dwarfs $\epsilon$~Indi~BC \citep{mcc04}]
are coeval with their primary stellar companions, constraints can be placed on the often cited -- but empirically untested -- evolutionary 
models of brown dwarfs \citep{bur97,cha00}.  

The estimated mass ratios of the four binary M dwarfs are evenly distributed into low ($\lesssim 0.25$) and high ($\gtrsim 0.75$) bins.  
We can hardly draw statistical inferences from such a small sample, but it is interesting that the first identifications of binary M dwarfs
from our NICMOS survey have a bimodal distribution of mass ratios.  Such a distribution defies the noted tendency for binary M dwarfs to have 
components of approximately equal mass \citep{rei97a,rei97b}.  On the other hand, the nearly identical components of GJ~1001BC conform to 
the strong trend toward equal masses observed among binary L and T dwarfs \citep{rei01,bur03,bou03}.  We defer to a later paper a complete
discussion of the systemic statistics (i.e., multiplicity fraction, secondary luminosity function, and distribution of mass ratios) of the
RECONS ``10~pc sample'' once our NICMOS survey is complete and fully analyzed.

\acknowledgments 
The authors thank Kuenley Chiu and Wei Zheng for obtaining the SPICam images of G239--25AB on our behalf.  We also thank Mark Dickinson and
Russet McMillan for their advice concerning the calibration of our NICMOS and SPICam data, respectively.  Support for the NICMOS snapshot 
survey was provided by NASA through grants HST-GO-07420, HST-GO-07894, and HST-GO-09485 from the Space Telescope Science Institute (STScI).
The FGS studies have been supported by NASA through grants HST-GO-08532, HST-GO-08729, HST-GO-09408, and HST-GO-9972 from the STScI.  ACS 
was developed under NASA contract NAS5-32865, and ACS IDT research has been supported by NASA grant NAG5-7697 and by an equipment grant from
Sun Microsystems, Inc.  The STScI is operated by AURA, Inc., under NASA contract NAS5-26555.\\

\noindent
{\it Note added in proof --}  Resolved images of LHS~224AB and GJ~84AB have been published contemporaneously in {\it Astronomy \& Astrophysics} by 
J.-L.\ Beuzit et al.

\newpage 

\begin{deluxetable}{llrrccclrrrccccc}
\rotate
\tabletypesize{\scriptsize}
\tablenum{1}
\tablewidth{580pt}
\tablecaption{Dwarfs with newly discovered or confirmed companions}
\tablehead{
 & \colhead{RA} & \colhead{Dec} & \colhead{$b$} & \colhead{$\pi$ (error)\tablenotemark{a}} & \multicolumn{2}{c}{Proper motion} & & \multicolumn{3}{c}{Photometry\tablenotemark{b}}& & \multicolumn{4}{c}{References\tablenotemark{c}} \\
\cline{2-3}\cline{6-7}\cline{9-11}\cline{13-16}
\colhead{Name} & \multicolumn{2}{c}{(J2000)} & \colhead{($^\circ$)} & \colhead{(mas)} & \colhead{$\mu$ ($''$ yr$^{-1}$)} & \colhead{PA ($^\circ$)} & \colhead{SpT} & $K_s$ & $V$--$K_s$ & $J$--$K_s$ & & \colhead{Coords.} & \colhead{$\pi$} & \colhead{$\mu$} & \colhead{SpT}
}
\tablecolumns{16}
\startdata 
GJ 1001A       & 00:04:36.4 & $-40$:44:03 & $-73$ & 104.7~ (11.4~) & 1.618 & 154.5 & M3.5 &  7.74 &    5.10 & 0.86 & & 1 & 2 & 3 & 4 \\
GJ 1001B       & 00:04:34.9 & $-40$:44:06 & $-73$ & 104.7~ (11.4~) & 1.618 & 154.5 & L4.5 & 11.40 & \nodata & 1.71 & & 5 & 2 & 3 & 6 \\
GJ~54	       & 01:10:22.9 & $-67$:26:42 & $-50$ & 122.02 (~6.04) & 0.692 & ~34.2 & M3.0 &  5.13 &    4.67 & 0.88 & & 7 & 2,7 & 7 & 8 \\
GJ~84	       & 02:05:04.8 & $-17$:36:53 & $-71$ & 106.37 (~1.98) & 1.329 & ~97.5 & M3.0 &  5.66 &    4.55 & 0.81 & & 7 & 2,7 & 7 & 8 \\
LHS 224        & 07:03:55.9 & $+52$:42:06 & $+23$ & 108.5~ (~2.1~) & 1.166 & 141.7 & M5.0 &  7.78 &    5.51 & 0.76 & & 1 & 2 & 3 & 9 \\
G~239--25      & 14:42:21.6 & $+66$:03:21 & $+47$ & 101.34 (12.77) & 0.316 & 259.2 & M3.0 &  6.49 &    4.34 & 0.81 & & 7 & 7 & 7 & 3 \\
\enddata

\tablenotetext{a}{\scriptsize Based on weighted mean of referenced trigonometric parallaxes.}
\tablenotetext{b}{\scriptsize $J$ and $K_s$ from Two-Micron All Sky Survey (2MASS); $V$ from \citet{gli91} or \citet{wei88,wei96}.}
\tablenotetext{c}{\scriptsize References for coordinates, trigonometric parallax, proper motion, and spectral type: 
(1) \citealt{bak02};
(2) \citealt{van95};
(3) \citealt{gli91};
(4) \citealt{mar99};
(5) 2MASS;
(6) \citealt{leg02b};
(7) \citealt{esa97};
(8) T.~Beaulieu et al., in preparation;
(9) \citealt{rei95} 
} 
\end{deluxetable}

\newpage

\begin{deluxetable}{lcccccc}
\tabletypesize{\footnotesize}
\tablenum{2}
\tablewidth{480pt}
\tablecaption{Log of observations}
\tablehead{
\colhead{Item}      & \colhead{GJ 1001A}  & \colhead{GJ 1001B}  & \colhead{GJ~54} & \colhead{GJ~84} & \colhead{LHS 224} & \colhead{G~239--25}
}
\tablecolumns{7}
\startdata
Epoch 1 \\
\hspace*{0.15in} Date (UT)  & 1998 Aug 03 & 2002 Oct 11 & 1998 Nov 09 & 2002 Oct 02 & 2003 Mar 13 & 1998 Nov 07 \\
\hspace*{0.15in} Telescope  & {\it HST}   & {\it HST}   & {\it HST}   & {\it HST}   & {\it HST}   & {\it HST}   \\
\hspace*{0.15in} Instrument & NIC2        & NIC2        & NIC2        & NIC2        & NIC2        & NIC2        \\
\hspace*{0.15in} Program No. & 7894        & 9485        & 7894        & 9485        & 9485        & 7894        \\
Epoch 2 \\
\hspace*{0.15in} Date (UT)  & 2003 Aug 20 & 2003 Aug 20 & 2000 Sep 24 & \nodata     & 2000 Jan 07\tablenotemark{a} & 2003 May 06 \\
\hspace*{0.15in} Telescope  & {\it HST}   & {\it HST}   & {\it HST}   & \nodata     & {\it HST}   & ARC 3.5~m   \\
\hspace*{0.15in} Instrument & ACS/HRC     & ACS/HRC     & FGS\tablenotemark{b}         & \nodata     & FGS\tablenotemark{b}         & SPICam      \\
\hspace*{0.15in} Program No. & 9990        & 9990        & 8729        & \nodata     & 8532        & \nodata     \\
\enddata
\tablenotetext{a}{\footnotesize Observed in independent FGS search for multiplicity among nearby M dwarfs and white dwarfs.}
\tablenotetext{b}{\footnotesize Monitoring of orbit continues as part of {\it HST} Programs 9408, 9972, and 10104.}
\end{deluxetable}

\newpage

\begin{deluxetable}{lcccc}
\tabletypesize{\small}
\tablenum{3}
\tablewidth{445pt}
\tablecaption{NIC2 exposure characteristics}
\tablehead{
\colhead{Item}          & \colhead{F110W}  & \colhead{F180M} & \colhead{F207M} & \colhead{F222M}
}
\tablecolumns{5}
\startdata
Mean $\lambda$ ($\mu$m)                             & 1.1035     & 1.7971      & 2.0786      & 2.2164      \\
FWHM ($\mu$m)                                       & 0.5915     & 0.0684      & 0.1522      & 0.1432      \\
MULTIACCUM sampling sequence\tablenotemark{a}	    & STEP64     & STEP64      & STEP128     & STEP128     \\
$PHOTFNU$ (Jy~s~DN$^{-1}$; 62~K)\tablenotemark{b}   & 1.86394e-6 & 1.07637e-05 & 5.38945e-06 & 5.11497e-06 \\
$PHOTFNU$ (Jy~s~DN$^{-1}$; 77.1~K)\tablenotemark{c} & 1.2128~e-6 & 7.9701~e-06 & 4.2433~e-06 & 4.1897~e-06 \\
$F_{\nu}$(Vega) (Jy)                  		    & 1898       & 932         & 735         & 653         \\
\enddata

\tablenotetext{a}{STEP$n$ = 0.3, 0.6, 1, 2, 4, 8, \ldots , $n$ (seconds).} 
\tablenotetext{b}{Values before installation of NCS. (M.~Dickinson 2004, private communication)}
\tablenotetext{c}{Values after installation of NCS. (M.~Dickinson 2004, private communication)}
\end{deluxetable}

\newpage
 
\begin{deluxetable}{lrrrrcccc}
\rotate
\tabletypesize{\footnotesize}
\tablenum{4}
\tablewidth{580pt}
\tablecaption{NICMOS photometry and astrometry of multiple systems}
\tablehead{
               & \multicolumn{4}{c}{Apparent magnitude (error)\tablenotemark{a}}         &                   & \multicolumn{2}{c}{Separation (error)}               & \colhead{PA (error)\tablenotemark{c}}\\
\cline{2-5}\cline{7-8}
\colhead{Name} & \colhead{F110W}  & \colhead{F180M}  & \colhead{F207M} & \colhead{F222M} & \colhead{Method\tablenotemark{b}} & \colhead{Angular} & \colhead{Projected (AU)} & \colhead{(deg)} 
}
\tablecolumns{9}
\startdata
GJ~1001A       &  8.89 (0.04)   &  7.97 (0.04)   &  7.86 (0.04)  &  7.72 (0.04)  & Aperture  & \nodata        & \nodata        & \nodata \\
GJ~1001B       & 14.50 (0.09)   & 12.86 (0.05)   & 12.47 (0.04)  & 12.04 (0.04)  & PSF fit   & \raisebox{-1.5ex}{0\farcs087 (0\farcs006)}  & \raisebox{-1.5ex}{~0.83 (0.11)} & \raisebox{-1.5ex}{~48.1 (5.2)\tablenotemark{d}} \\
\raisebox{0.5ex}{GJ~1001C}       & \raisebox{0.5ex}{14.60 (0.09)}   & \raisebox{0.5ex}{13.00 (0.05)}   & \raisebox{0.5ex}{12.31 (0.04)}  & \raisebox{0.5ex}{12.05 (0.04)}  & \raisebox{0.5ex}{PSF fit} \\
 & \\
GJ~54A         & 6.72 (0.12)    &  5.80 (0.05)   &  5.73 (0.05)  &  5.65 (0.05)  & PSF fit   & \raisebox{-1.5ex}{0\farcs129 (0\farcs006)}       & \raisebox{-1.5ex}{~1.05 (0.07)}       & \raisebox{-1.5ex}{284.9 (3.8)} \\ 
\raisebox{0.5ex}{GJ~54B}        & \raisebox{0.5ex}{7.28 (0.12)}       &  \raisebox{0.5ex}{6.59 (0.05)}  &  \raisebox{0.5ex}{6.51 (0.05)} &  \raisebox{0.5ex}{6.37 (0.05)} & \raisebox{0.5ex}{PSF fit} \\
 & \\
GJ~84A         &  6.38 (0.12)   &  5.75 (0.05)   &  5.55 (0.04)  &  5.57 (0.04)  & PSF fit   & \raisebox{-1.5ex}{0\farcs443 (0\farcs006)}       & \raisebox{-1.5ex}{~4.17 (0.10)}       & \raisebox{-1.5ex}{103.4 (1.0)} \\ 
\raisebox{0.5ex}{GJ~84B}         & \raisebox{0.5ex}{10.97 (0.09)}  &  \raisebox{0.5ex}{9.76 (0.05)}  &  \raisebox{0.5ex}{9.73 (0.04)} &  \raisebox{0.5ex}{9.39 (0.04)} & \raisebox{0.5ex}{PSF fit} \\
 & \\
LHS~224A       &  9.35 (0.12)   &  8.71 (0.09)   &  8.59 (0.09)  &  8.37 (0.09)  & PSF fit   & \raisebox{-1.5ex}{0\farcs132 (0\farcs012)}       & \raisebox{-1.5ex}{~1.21 (0.11)}       & \raisebox{-1.5ex}{~14.0 (7.2)} \\ 
\raisebox{0.5ex}{LHS~224B}       &  \raisebox{0.5ex}{9.36 (0.12)}  &  \raisebox{0.5ex}{8.88 (0.09)}  &  \raisebox{0.5ex}{8.75 (0.09)} &  \raisebox{0.5ex}{8.51 (0.09)} & \raisebox{0.5ex}{PSF fit} \\
 & \\
G~239--25A     &  7.66 (0.09)   &  6.74 (0.05)   &  6.68 (0.05)  &  6.59 (0.05)  & PSF fit   & \raisebox{-1.5ex}{3\farcs041 (0\farcs006)}       & \raisebox{-1.5ex}{30.01 (3.78)}       & \raisebox{-1.5ex}{114.0 (0.2)} \\ 
\raisebox{0.5ex}{G~239--25B}     & \raisebox{0.5ex}{11.93 (0.04)}  & \raisebox{0.5ex}{10.83 (0.04)}  & \raisebox{0.5ex}{10.58 (0.04)} & \raisebox{0.5ex}{10.28 (0.04)} & \raisebox{0.5ex}{Aperture} \\
\enddata

\tablenotetext{a}{Errors include uncertainties in the following quantities, combined in quadrature (where applicable): photon noise, read noise, 
unsaturated PSF fits (3--8\%), PSF fits affected by saturation or bad pixels (8\%), absolute photometric calibration ($< 3$\%), and zero point 
drift ($< 2$\%).}
\tablenotetext{b}{Method for determining integrated count rates (see \S2.1).}
\tablenotetext{c}{Position angle measured east of north.}
\tablenotetext{d}{Photometric uncertainties make the identities of GJ~1001B and C ambiguous.  Consequently, the PA suffers an ambiguity of
$180^{\circ}$.}
\end{deluxetable}

\newpage

\begin{deluxetable}{lcccc}
\tablenum{5}
\tablewidth{345pt}
\tablecaption{FGS photometry and astrometry of binary M dwarfs\tablenotemark{a}}
\tablehead{
\colhead{Name} & \colhead{$\Delta$F583W (error)} & \colhead{Separation} & \colhead{PA (deg)} & \colhead{Epoch}
}
\tablecolumns{5}
\startdata
GJ~54AB        & 1.040 (0.074)                   & 0\farcs11      & ~~93               & 2000.73 \\
               &                                 & 0\farcs13      & 283               & 2003.45 \\
LHS~224AB      & 0.293 (0.006)                   & 0\farcs15      & ~~~5               & 2000.02 \\
               &                                 & 0\farcs16      & 330               & 2003.83 \\
\enddata

\tablenotetext{a}{Preliminary measurements of limited precision are given pending completion of the
FGS monitoring programs.  (See \S2.2.1.)}
\end{deluxetable}

\newpage

\begin{deluxetable}{lrrrcrrc}
\rotate
\tabletypesize{\footnotesize}
\tablenum{6}
\tablewidth{530pt}
\tablecaption{ACS photometry and astrometry of the GJ~1001 system}
\tablehead{
               & \multicolumn{3}{c}{Apparent magnitude\tablenotemark{a} ~(error)\tablenotemark{b}}         &                   & \multicolumn{2}{c}{Separation (error)} & \colhead{PA (error)\tablenotemark{d}}\\
\cline{2-4}\cline{6-7}
\colhead{Name} & \colhead{F625W}  & \colhead{F775W}  & \colhead{F850LP} & \colhead{Method\tablenotemark{c}} & \colhead{Angular} & \colhead{
Projected (AU)} & \colhead{(deg)}
}
\tablecolumns{8}
\startdata
GJ~1001A & 12.11 (0.02) & 10.67 (0.02) &  9.94 (0.02) & Aperture & \raisebox{-1.5ex}{18\farcs236 (0\farcs002)\tablenotemark{e}} & \raisebox{-1.5ex}{174.17 (18.96)\tablenotemark{e}} & \raisebox{-1.5ex}{258.8 ($\ll 0.1$)\tablenotemark{e}} \\
\raisebox{0.5ex}{GJ~1001B} & \raisebox{0.5ex}{20.80 (0.04)} & \raisebox{0.5ex}{18.35 (0.07)} & \raisebox{0.5ex}{16.42 (0.11)} & \raisebox{0.5ex}{PSF fit} & \raisebox{-1.0ex}{0\farcs086 (0\farcs002)} & \raisebox{-1.0ex}{0.82~ (0.09)} & \raisebox{-1.0ex}{95.5 (0.5)} \\
\raisebox{0.5ex}{GJ~1001C} & \raisebox{0.5ex}{20.93 (0.04)} & \raisebox{0.5ex}{18.50 (0.07)} & \raisebox{0.5ex}{16.52 (0.11)} & \raisebox{0.5ex}{PSF fit} \\
\enddata
\tablenotetext{a}{AB magnitudes are given to permit comparison with SDSS $r$, $i$, and $z$ photometry.  To obtain Vega-based F625W, F775W, 
and F850LP magnitudes, subtract 0.167, 0.398, and 0.536, respectively, from the AB magnitudes listed in the table, and convolve an 
additional error of 0.02~mag to account for the uncertainty in the absolute flux calibration of Vega \citep{sir04}.}
\tablenotetext{b}{Photometric errors include uncertainties in the following quantities, combined in quadrature (where applicable): photon noise, 
read noise, PSF fitting (3--10\%), flat field (1\%), instrument stability (0.5\%), and charge-transfer inefficiency (1.5\%).}
\tablenotetext{c}{Method for determining integrated count rates (see \S2.2.3).}
\tablenotetext{d}{Position angle measured east of north.}
\tablenotetext{e}{Separation and PA of GJ~1001BC's photocenter with respect to GJ~1001A.}
\end{deluxetable}

\newpage

\begin{deluxetable}{llccccccc}
\rotate
\tabletypesize{\footnotesize}
\tablenum{7}
\tablewidth{520pt}
\tablecaption{Known binary L and T dwarfs with periods $\lesssim 15$~yr}
\tablehead{
               &               & \colhead{Dist.} & \multicolumn{2}{c}{Separation (error)}       & \colhead{Period\tablenotemark{a}} & \multicolumn{3}{c}{References\tablenotemark{b}}\\
\cline{4-5}\cline{7-9}
\colhead{Name} & \colhead{SpT} & \colhead{(pc)}                   & \colhead{Ang.} & \colhead{Proj.\ (AU)} & \colhead{(yr)}                    & \colhead{Image}  & \colhead{SpT} & \colhead{$\pi$}
}
\tablecolumns{8}
\startdata
GJ~1001BC                  & L4.5 + L4.5        & 9.6--15               & 0\farcs087           & 0.8--1.3  & 3--5.5     & 1 & 2 & 3,4 \\
2MASS~J15344984--2952274AB & T5.5 + T5.5        & 13.6                  & 0\farcs065           & 0.9       & 3--5       & 5 & 5 & 6 \\
2MASSW~J0920122+351742AB   & L6.5 + $\sim$~L7.0 & 20.8\tablenotemark{c} & 0\farcs070           & 1.5       & 6--10      & 7 & 7 & 7 \\
GJ~417BC\tablenotemark{d}  & L4.5 + $\sim$~L5.5 & 21.7                  & 0\farcs070           & 1.5       & 7--11      & 8 & 9 & 10 \\
2MASS~J07464256+2000321AB  & L0.0 + L1.5        & 12.2                  & 0\farcs12--0\farcs22 & 1.5--2.7  & 8--13      & 7,11,12 & 12 & 13 \\
GJ~564BC		   & L4.0 + L4.0        & 17.9			& 0\farcs134	       & 2.4       & 14--22     & 14 & 15 & 10 \\
$\epsilon$~Indi~BC         & T1.0 + T6.0        & 3.6                   & 0\farcs732           & 2.6       & 15--25     & 16 & 16 & 10 \\
\enddata

\noindent
\tablenotetext{a}{Periods of all binaries except 2MASS~J0746+2000AB are estimated from equation~(2) and assumed masses of 0.030--$0.075~M_{\odot}$
(0.060--$0.075~M_{\odot}$ for GJ~1001BC; see text).  2MASS~J0746+2000AB's period was computed by \citet{bou04} from a Keplerian fit to 60\% of 
its orbit.}
\tablenotetext{b}{References for resolved images, spectral types, and parallax: (1) this paper; (2) \citealt{leg02b}; (3) \citealt{van95}; 
(4) T.~Henry et al., in preparation; (5) \citealt{bur03}; (6) \citealt{tin03}; (7) \citealt{rei01}; (8) \citealt{bou03}; (9) \citealt{kir00}; 
(10) \citealt{esa97}; (11) \citealt{clo03}; (12) \citealt{bou04}; (13) \citealt{dah02}; (14) \citealt{pot02}; (15) \citealt{got02}; 
(16) \citealt{mcc04}.}
\tablenotetext{c}{Estimated from photometric parallax.}
\tablenotetext{d}{Also known as 2MASS~J1112257+354813AB.}
\end{deluxetable}

\newpage

\newpage

\newpage

\newpage

\newpage

\newpage

\newpage

\end{document}